\documentclass[journal]{IEEEtran}

\usepackage{comment}
\usepackage[T1]{fontenc}
\usepackage{algorithmic}
\usepackage{algorithm}
\usepackage{color}
\usepackage{graphicx}

\usepackage{mathrsfs}
\usepackage{amsfonts,amssymb}
\usepackage{subfigure}
\usepackage{amsmath}
\usepackage[justification=centering]{caption}
\ifCLASSINFOpdf
\else
\fi

\usepackage{multirow}
\usepackage{amsmath}
\usepackage{bm}
\usepackage[cmintegrals]{newtxmath}
\usepackage{xcolor}
\usepackage{xpatch}

\makeatletter
\def\changeBibColor#1{%
	\in@{#1}{pan2017joint, 8438896}	
	\ifin@\color{black}\else\normalcolor\fi
}

\xpatchcmd\@bibitem
{\item}
{\changeBibColor{#1}\item}
{}{\fail}

\xpatchcmd\@lbibitem
{\item}
{\changeBibColor{#2}\item}
{}{\fail}
\makeatother

\begin{document}
\title{Deep Reinforcement Learning Based Dynamic Trajectory Control for UAV-assisted Mobile Edge Computing}

\author{Liang Wang,
        Kezhi Wang,
        Cunhua Pan,
        Wei Xu,
        Nauman Aslam and Arumugam Nallanathan, \textsl{Fellow, IEEE}

\thanks{
(Corresponding Author: Kezhi Wang)
	
This work of W. Xu was supported in part by the NSFC under grants 62022026 and 61871109.
	
	Liang, Kezhi and Nauman are with the Department
of Computer and Informantion Science, Northumbria University, Newcastle upon Tyne, UK, NE1 8ST, emails: \{liang.wang, kezhi.wang, nauman.aslam\}@northumbria.ac.uk.

Cunhua and Arumugam are with School
of Electronic Engineering and Computer Science, Queen Mary University of London, E1 4NS, U.K., emails: \{c.pan, a.nallanathan\}@qmul.ac.uk

W. Xu is with the National Mobile Communications Research Lab, Southeast University, Nanjing 210096, China, and also with Henan Joint International Research Laboratory of Intelligent Networking and Data Analysis, Zhengzhou University, Zhengzhou, 450001 China (wxu@seu.edu.cn).
}

}

\maketitle
\begin{abstract}
In this paper, we consider a platform of flying mobile edge computing (F-MEC), where unmanned aerial vehicles (UAVs) serve as equipment providing computation resource, and they enable task offloading from user equipment (UE). We aim to minimize energy consumption of all UEs via optimizing user association, resource allocation and the trajectory of UAVs. To this end, we first propose a Convex optimizAtion based Trajectory control algorithm (CAT), which solves the problem in an iterative way by using block coordinate descent (BCD) method. Then, to make the real-time decision while taking into account the dynamics of the environment (i.e., UAV may take off from different locations), we propose a deep Reinforcement leArning based Trajectory control algorithm (RAT). In RAT, we apply the Prioritized Experience Replay (PER) to improve the convergence of the training procedure. Different from the convex optimization based algorithm which may be susceptible to the initial points and requires iterations, RAT can be adapted to any taking off points of the UAVs and can obtain the solution more rapidly than CAT once training process has been completed. Simulation results show that the proposed CAT and RAT achieve the considerable performance and both outperform traditional algorithms.

\end{abstract}


\IEEEpeerreviewmaketitle

\section{Introduction} 
\IEEEPARstart{W}{ith} the popularity of computationally-intensive tasks, e.g., smart navigation and augmented reality, people are expecting to enjoy more convenient life than ever before. However, current smart devices and user equipments (UEs), due to small size and limited resource, e.g., computation and battery, may not be able to provide satisfactory Quality of Service (QoS) and Quality of Experience (QoE) in executing those highly demanding tasks. 

Mobile edge computing (MEC) has been proposed by moving the computation resource to the network edge and it has been proved to greatly enhance UE's ability in executing computation-hungry tasks \cite{hu2015mobile}. Recently, flying mobile edge computing (F-MEC) has been proposed, which goes one step further by considering that the computing resource can be carried by unmanned aerial vehicles (UAVs) \cite{8647789}. F-MEC inherits the merits of UAV and it is expected to provide more flexible, easier and faster computing service than traditional fixed-location MEC infrastructures. However, the F-MEC also brings several challenges: 1) how to minimize the long-term energy consumption of all UEs by choosing proper user association (i.e., whether UE should offload the tasks and if so, which UAV to offload to, in the case of multiple flying UAVs); 2) how much computations the UAV should allocate to each offloaded UE by considering the limited amount of on-board resource; 3) how to control each UAV's trajectory in real time (namely, flying direction and distance), especially considering the dynamic environment (i.e., the UAV may serve UEs from different taking off points). Traditional approaches like exhaustive search are hardly to tackle the above problems due to the fact that the decision variable space of F-MEC, e.g., deciding the optimal trajectory and resource allocation, is continuous instead of discrete. In \cite{8274943}, the authors propose a quantized dynamic programming algorithm to address the resource allocation problem of MEC. However, the complexity of this approach is very high as the flying choice of UAV is nearly infinite (as continues variables). Moreover, the authors in \cite{8438896} discretize the UAV trajectory into a sequence of UAV locations and make their proposed problem tractable. Similarly, in \cite{8637952}, the authors assume that the UAV's trajectory can be approximated by using the discrete variables and then they deal with it by using the traditional convex optimization approaches. However, the above treatment may decrease the control accuracy of the UAV and also is not flexible. Furthermore, the above contributions only considered a single UAV case. In practice, one UAV may not have enough resource to serve all the users. If the served area is very large, more than one UAV are normally needed, which will undoubtedly increase the decision space and make it very difficult for the traditional convex optimization-based approaches to obtain the optimal control strategies of each UAV. In~\cite{8432464}, Liu \emph{et al}. propose a deep reinforcement learning based DRL-EC$^3$ algorithm, which can control the trajectory of multiple UAVs but did not consider the user association and resource allocation. 

Inspired by the challenges mentioned above, in this paper, we first propose a Convex optimizAtion based Trajectory control algorithm (CAT) to minimize the energy consumption of all the UEs, by jointly optimizing user association, resource allocation and UAV trajectory. Specifically, by applying block coordinate descent (BCD) method, CAT is divided into two parts, i.e., subproblems for deciding UAV trajectories and for deciding user association and resource allocation. In each iteration, we solve each part separately while keep the other part fixed, until the convergence is achieved.

Next, we propose a deep Reinforcement leArning based Trajectory control algorithm (RAT) to facilitate the real-time decision making. In RAT, two deep Q networks (DQNs), i.e., actor and critic networks are applied, where the actor network is responsible for deciding the direction and flying distance of the UAV, while the critic network is in charge of evaluating the actions generated by the actor network. Then, we propose a low-complexity matching algorithm to decide the user association and resource allocation with the UAVs. 
We choose the overall energy consumption of all the UEs as a reward of the RAT. In addition, we deploy a mini-batch to collect samples from the experience replay buffer by using a Prioritized Experience Replay (PER) scheme. 

Different from traditional optimization based algorithms which normally need iterations and are susceptible to initial points, the proposed RAT can be adapted to any taking off points of the UAVs and can obtain the solutions very rapidly once the training process has been completed. In other words, if the taking off points of UAV are input to the RAT, the trajectories of UAVs will be determined by the proposed RAT with only some simple algebraic calculations instead of solving the original optimization problem through traditional high-complexity optimization algorithms. This attributes to the fact that during the training stages, excessive randomly taking off points of UAV are generated and used to train the networks until they are converged. Also, with the help of prioritized experience reply (PER), the convergence speed will be increased significantly. RAT can be applied to the practical scenarios where the UAVs needs to act and fly swiftly such as the battlefields. By inputting the current coordinates as the taking off points to the networks, the trajectories of the UAVs will be immediately obtained and then all the UAVs can take off and fly according to the obtained trajectories. Also, the resource allocation and user association are determined by the proposed low-complexity matching algorithm. This is particularly useful to some emergence scenarios (e.g., battlefields, earthquake, large fires), as fast decision making is crucial in these areas.

In the simulation, we can see that the proposed RAT can achieve the similar performance as the convex-based solution CAT. They both have considerable performance gain over other traditional algorithms. In addition, we can see that during the learning procedure, the proposed RAT is less sensitive to the hyperparameters, i.e., the size of mini-batch and the experience replay buffer, when comparing to tradtional reinforcement learning where PER is not applied.

The remainder of this paper is organized as follows. Section \uppercase\expandafter{\romannumeral2} presents the related work. 
Section \uppercase\expandafter{\romannumeral3} describes the system model. Section \uppercase\expandafter{\romannumeral4} introduces the proposed CAT algorithm, whereas Section \uppercase\expandafter{\romannumeral5} gives the proposed RAT algorithm including the preliminaries of DRL. Section \uppercase\expandafter{\romannumeral6} extends the application of proposed RAT algorithm to 3-D scenario. The simulation results are reported in Section \uppercase\expandafter{\romannumeral7}. Finally, conclusions are given in Section \uppercase\expandafter{\romannumeral8}.

\section{Related Work}
There are many related works that study UAV, MEC and DRL separately, but only a very few consider them holistically. For UAV aided wireless communications, several scenarios have been studied, such as in areas of  relay transmissions~\cite{kong2017autonomous}, cellular system~\cite{challita2019machine}, data collection~\cite{8119562}, wireless power transfer~\cite{8365881}, caching networks~\cite{zhao2018caching}, and D2D communication~\cite{mozaffari2016unmanned}. In~\cite{al2014optimal}, the authors presented an approach to optimize the altitude of UAV to guarantee the maximum radio coverage on the ground. In~\cite{8103781}, the authors presented a fly-hover-and-communicate protocol in a UAV-enabled multiuser communication system. They partitioned the ground terminals into disjoint clusters and deployed the UAV as a flying base station. Then, by jointly optimizing the UAV altitude and antenna beamwidth, they optimized the throughput in UAV-enabled downlink multicasting, downlink broadcasting, and uplink multiple access models. In~\cite{8438896}, to maximize the minimum average throughput of covered users in OFDMA system, the authors proposed an efficient iterative algorithm based on block coordinate descent and convex optimization techniques to optimize the UAV trajectory and resource allocation. Furthermore, UAV trajectory optimization research were also investigated. For instance in~\cite{7888557}, Zeng~\emph{et al.} proposed an efficient design by optimizing UAV's flight radius and speed for the sake of maximizing the energy efficiency of UAV communication. In order to maximize the minimum throughput of all mobile terminals in cellular networks, Lyu~\emph{et al.}~\cite{lyu2018uav} developed a new hybrid network architecture by deploying UAV as an aerial mobile base station. Different from~\cite{al2014optimal, 8103781, 8438896, 7888557} with the single UAV system, a multi-UAV enabled wireless communication system was considered to serve a group of users in~\cite{8247211}. Also, in~\cite{Yang_2019}, resource allocation between communication and computation has been investigated in multi-UAV systems. \textcolor{black}{In~\cite{8038869}, Mozaffari~\emph{et al.} investigated the application of UAVs in Internet of Things (IoT) network, and they  optimized the mobility of UAVs, the device-UAV association and uplink power control, for minimizing the overall transmit power of ground IoT devices.}

In addition, some recent literature made efforts to mobile edge computing (MEC), which is considered to be a promising technology for bringing computing resource to the edge of wireless networks~\cite{mao2017survey}, where UEs can benefit from offloading their tasks to MEC servers. In~\cite{7929399}, partial computation offloading was studied. The computation tasks can be divided into two parts, where one part is executed locally and the other part is offloaded to MEC servers. In~\cite{6574874}, binary computation offloading was studied, where the computation tasks can either be executed locally or offloaded to MEC servers.

By taking advantage of the mobility of UAVs, UAV-enabled MEC has been studied in~\cite{9275621, 8805125}. In~\cite{9275621}, authors proposed a heterogeneous MEC (H-MEC) architecture that consists of fixed ground stations and UAVs. In~\cite{8805125}, the authors studied UAV-enabled MEC, where wireless power transfer technology is applied to power Internet of things devices and collect data from them. \textcolor{black}{In~\cite{zhou2018computation}, Zhou~\emph{et al.} investigated an UAV-enabled MEC wireless-powered system, and they tackled the computation maximization problem through optimizing UAV's speed, partial and binary computation offloading modes.} \textcolor{black}{In~\cite{8740949}, Asheralieva~\emph{et al.} studied network operation problem in UAV-enabled MEC network, and they developed a framework based on hierarchical game-theoretic and reinforcement learning.} \textcolor{black}{In~\cite{8952621}, Zhang~\emph{et al.} established a communication and computation optimization model in an MEC-enabled UAV network, where the successful transmission probability was derived through using stochastic geometry.}

For most of the above works, optimization theory are mainly applied in order to obtain the optimal and / or suboptimal solutions, e.g., trajectory design and resource allocation. However, solving such optimization problems normally requires plenty of computational resources and take much time. To address this problem, DRL has been applied and attracted much attention recently. In~\cite{mnih2015human}, the authors proposed a RL framework that uses DQN as the function approximator. In addition, two important ingredients experience replay and target network are used for improving the convergence performance. In~\cite{hasselt2015deep}, the authors pointed out that the classical DQN algorithm may suffer from substantial overestimations in some scenarios, and proposed a double Q-learning algorithm. In order to solve control problems with continuous state and action space, Lillicrap \emph{at al.}~\cite{lillicrap2015continuous} proposed a policy gradient based algorithm. For the purpose of obtaining faster learning and state-of-art performance, in~\cite{schaul2015prioritized}, the authors proposed a more robust and scalable approach named prioritized experience replay. Although DRL has achieved remarkable successes in game-playing scenarios, it is still an open research area in UAV-enabled MEC. 

\section{System Model}\label{section2}

\begin{figure}[htpb]
	\centering
	\includegraphics[width=3.6in]{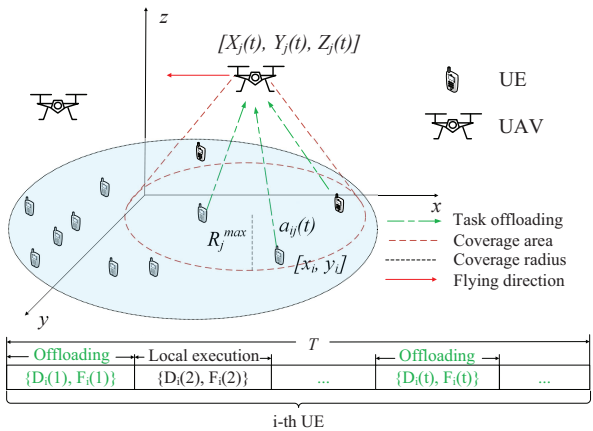}
	\caption{Multi-UAV enabled F-MEC architecture.}\label{fig1}
\end{figure}

As shown in Fig.~\ref{fig1}, we consider a scenario that there are $N$ UEs with the set denoted as $ \mathcal{N} = \{1, 2,...,N\}$ and $M$ UAVs with the set denoted as $ \mathcal{M} = \{1,2,...,M\}$, which form an F-MEC platform. To make it clear, the main notations used in this paper are listed in Table.~\ref{nota}.

\begin{table}[htbp!]
	\begin{center}
		\caption{Main Notations.}
		\begin{tabular}{|c|c|c|c|}
			\hline
			\textbf{Notation} & \textbf{Definition} \\ \hline
			$i, N, \mathcal{N}$ & \text{index, number, set of of UEs.}  \\ \hline
			$j, M, \mathcal{M}$ & \text{index,number,set of UAVs.}   \\ \hline
			$t, T, \mathcal{T}$ & \text{index, number, set of time slots.}   \\ \hline
			$I_i(t),D_i(t), F_i(t)$ & \text{$i$-th UEs' task in $t$-th time slot.}   \\ \hline
			$a_{ij}(t)$ & \text{user association between $i$-th UE and $j$-th UAV. }   \\ \hline
			$R^{\text{max}}_j$ & \text{maximal horizontal coverage radius of $j$-th UAV.}   \\ \hline
			$\theta^h_j(t), \theta^v_j(t), d_j(t)$ & \text{flying action of $j$-th UAV.}   \\ \hline
			$d^{\text{max}}, v_j(t)$ & \text{maximal distance, velocity of $j$-th UAV.}   \\ \hline
			$[X_j(t), Y_j(t), Z_j(t)]$ & \text{coordinate of $j$-th UAV.}   \\ \hline
			$X^{\text{max}}, Y^{\text{max}}$ & \text{side length of rectangle-shaped area.}   \\ \hline
			$T^{\text{max}}$ & \text{maximal time duration.}   \\ \hline
			$V^{\text{max}}, f^{\text{max}}$ & \text{maximal number of tasks, computation resource.}   \\ \hline
			$[x_i, y_i]$ & \text{coordinate of $i$-th UE.}   \\ \hline
			$R_{ij}(t)$ & \text{horizontal distance between UE and UAV.}   \\ \hline
			$B, P^{\text{Tr}}$ & \text{channel bandwidth, transmitting power.}   \\ \hline
			$g_0, \sigma^2$ & \text{channel power gain, noise power.}   \\ \hline
			$T_{ij}^{\text{O}}(t), T_{ij}^{\text{Tr}}(t),  T_{ij}^{\text{C}}(t)$ & \text{time for task completion, offloading, executing.}   \\ \hline
			$E_{ij}^{\text{Tr}}(t), E_{ij}^{\text{L}}(t)$ & \text{energy for offloading, local execution.}   \\ \hline
			$\bm{U}, \bm{G}$ & \text{set of UAV trajectory, UAV coordinates. }   \\ \hline
			$\bm{A}, \bm{F}$ & \text{set of user association, resource allocation. }   \\ \hline
			$s(t), a(t), z(t)$ & \text{state, action and reward.}   \\ \hline
			$\pi(\cdot), Q(\cdot), L(\cdot)$ & \text{policy function, Q function, loss function.}   \\ \hline
			$K, X$ & \text{size of mini-batch, experience replay buffer.}   \\ \hline
			$\phi, \delta, J$ & \text{network parameter, TD-error, policy gradient.}   \\ \hline
			$Z^{\text{min}}, Z^{\text{max}}$ & \text{minimal, maximal altitude value.}   \\ \hline
			$d_{ij}(t)$ & \text{distance between the $j$-th UAV and $i$-th UE.}   \\ \hline
		\end{tabular}
		\label{nota}
	\end{center}
\end{table}

\textcolor{black}{We assume that the $i$-th UE generates one task $I_{i}(t)$ in the $t$-th time slot, which has to be executed within a maximal time duration $T^{\text{max}}$, due to the QoS requirement. In this paper, we assume the entire process lasts for $T$ time slots. Thus, $T$ tasks will be generated for each UE and we have $t \in \mathcal{T} = \{1,2,...,T\}$ and}

\begin{equation}\label{w1}
\begin{aligned}
I_{i}(t)=\{D_{i}(t),~F_{i}(t)\},~\forall i \in \mathcal{N}, t \in \mathcal{T},
\end{aligned}
\end{equation}
where $D_i(t)$ denotes the size of data required to be transmitted to a UAV if the UE chooses to offload the task, and $F_i(t)$ denotes the total number of CPU cycles needed to execute this task. Assume that each UE can choose either to offload the task to one of the UAVs or execute the task locally. Then one can have
\begin{equation}\label{w2}
\begin{aligned}
a_{ij}(t) = \{0, 1\}, \forall i \in \mathcal{N}, j \in \mathcal{M}, t \in \mathcal{T},
\end{aligned}
\end{equation}
where $a_{ij}(t) = 1$, $j \neq 0$ implies that the $i$-th UE decides to offload the task to the $j$-th UAV in the $t$-th time slot, while $a_{ij}(t) = 1$, $j =0$ means that the $i$-th UE executes the task itself in the $t$-th time slot, and otherwise, $a_{ij}(t) = 0$. Define a new set $j \in \mathcal{M}' = \{0,1,2,..., M\}$ to represent the possible place where the tasks from UEs can be executed, where $j=0$ indicates that UE conducts its own task locally without offloading.

In addition, we assume that each  UE can only be served by at most one UAV or itself, and each task only has one place to execute. Then, it follows
\begin{equation}\label{w3}
\begin{aligned}
\sum_{j=0}^{M}a_{ij}(t)=1, \forall i \in \mathcal{N}, t \in \mathcal{T}.
\end{aligned}
\end{equation}

\textcolor{black}{Additionally, in this paper, the OFDM is applied, which means that each UAV can only accept $V^{\text{max}}$ tasks in each time slot, due to the number of limited sub-carriers. Thus, one has} 
\begin{equation}\label{w6}
\begin{aligned}
\sum_{i=1}^{N}a_{ij}(t) \leq V^{\text{max}}, \forall j \in \mathcal{M}, t \in \mathcal{T}.
\end{aligned}
\end{equation}

\subsection{UAV Movement}
\textcolor{black}{Assume that the $j$-th UAV flies at the altitude and it has a maximal horizontal coverage, which depends on the azimuth angle of antennas and the flying altitude \cite{8103781}. Also, assume that in the $t$-th time slot, the $j$-th UAV can fly with a horizontal direction as}
\textcolor{black}{\begin{equation}
		\begin{aligned}
			0 \leq \theta_{j}^h(t) \leq 2 \pi, \forall j \in \mathcal{M}, t \in \mathcal{T},
		\end{aligned}
\end{equation}}
and distance as
\textcolor{black}{\begin{equation}\label{wdis}
		\begin{aligned}
			0 \leq d_{j}(t) \leq d^{\text{max}}, \forall j \in \mathcal{M}, t \in \mathcal{T},
		\end{aligned}
\end{equation}}
\textcolor{black}{where $d^{\text{max}}$ is the maximal flying distance that the UAV can move in each time slot, due to the limited power budget. In our paper, we describe the UAV's movement based on the Cartesian Coordinate system. Thus, we denote the coordinate of the $j$-th UAV in the $t$-th time slot as $[X_{j}(t), Y_{j}(t), Z_j]$, where $X_{j}(t)=X_{j}(0)+\sum_{l=1}^{t}d_{j}(l)\text{cos}\big(\theta_j^h(l)\big)$, $Y_{j}(t)=Y_{j}(0)+\sum_{l=1}^{t}d_{j}(l)\text{sin}\big(\theta_j^h(l)\big)$ and $[X_{j}(0), Y_{j}(0), Z_j]$ is the initial coordinate of the $j$-th UAV.}

\textcolor{black}{Additionally, each UAV can only move within a rectangle-shaped area, whose side length is denoted as $X^{\text{max}}$, and $Y^{\text{max}}$. Then, it has}

\textcolor{black}{\begin{equation}
		\begin{aligned}
			0 \leq X_j(t) \leq X^{\text{max}},~\forall j \in \mathcal{M}, t \in \mathcal{T},
		\end{aligned}
\end{equation}}
and
\textcolor{black}{\begin{equation}
		\begin{aligned}
			0 \leq Y_j(t) \leq Y^{\text{max}},~\forall j \in \mathcal{M}, t \in \mathcal{T}.
		\end{aligned}
\end{equation}}

\textcolor{black}{We denote that the $j$-th UAV can move with a constant velocity $v_j(t)$, which varies with the flying distance $d_j(t)$ in each time slot. Thus, it has}
\textcolor{black}{\begin{equation}
		\begin{aligned}
			v_j(t) = \frac{d_j(t)}{T^{\text{max}}},~\forall j \in \mathcal{M}, t \in \mathcal{T}.
		\end{aligned}
\end{equation}}

In this paper, we ignore the communication related energy, including communication circuitry and signal processing.

\subsection{Task Execution}
If the $i$-th UE decides to offload the task to the $j$-th UAV in the $t$-th time slot, then the horizontal distance $R_{ij}(t)$ can be written as
\begin{equation}\label{w7}
R_{ij}(t) = \sqrt{(X_{j}(t)-x_{i})^2+(Y_{j}(t)-y_{i})^2},
\end{equation}
where $[x_i, y_i]$ is the coordinate of the $i$-th UE. \textcolor{black}{Additionally, we assume that each UAV has a maximal azimuth angle $\theta^{\text{max}}$ \footnote{We define the azimuth angle with respect to a 3-D reference axis, such as $x$ axis, $y$ axis, $z$ axis.}. Thus, in each time slot, the maximal horizontal coverage of the $j$-th UAV $R^{\text{max}}$ can be obtained as follows}
\textcolor{black}{\begin{equation}
		\begin{aligned}
			R^{\text{max}} = Z_j \text{tan}(\theta^{\text{max}}).
		\end{aligned}
	\end{equation}
}
Thus, it has
\textcolor{black}{\begin{equation}\label{ww8}
		\begin{aligned}
			a_{ij}(t)R_{ij}(t) \leq R^{\text{max}},~\forall i \in \mathcal{N}, j \in \mathcal{M}, t \in \mathcal{T}.
		\end{aligned}
\end{equation}}

\textcolor{black}{In this paper, the free space channel model is applied. Thus, the uplink data rate is given by}
\begin{equation}\label{w8}
\begin{aligned}
r_{ij}(t)=B\log_2 \Bigg(1+\frac{\alpha P^{\text{Tr}}}{Z_{j}^2+R_{ij}^2(t)} \Bigg),~\forall i \in \mathcal{N}, j \in \mathcal{M}, t \in \mathcal{T},
\end{aligned}
\end{equation}
\textcolor{black}{where $B$ is the bandwidth for each communication channel; $P^{\text{Tr}}$ is the transmitting power of the $i$-th UE; $\alpha$=$\frac{g_0G_0}{\sigma^2}$ with $G_0$ $\approx$ \text{2.2846}~\cite{Yang_2019}; $g_0$ is the channel power gain at the reference distance \text{1} $m$ and $\sigma^2$ is the noise power. Note that we consider each user applies orthogonal frequency division multiplexing (OFDM) channel and there is no interference among them.}

If the $i$-th UE decides to offload its task to the $j$-th UAV in the $t$-th time slot, the total task completion time is given by
\begin{equation}\label{w9}
\begin{aligned}
T_{ij}^\text{O}(t) = T_{ij}^{\text{Tr}}(t) + T_{ij}^\text{C}(t),~\forall  t \in \mathcal{T},
\end{aligned}
\end{equation}
where $T_{ij}^{\text{Tr}}(t)$ is the time to offload the data from the $i$-th UE to the $j$-th UAV in the $t$-th time slot, given by
\begin{equation}\label{w10}
\begin{aligned}
T_{ij}^{\text{Tr}}(t) = \frac{D_i(t)}{r_{ij}(t)},~\forall t \in \mathcal{T},
\end{aligned}
\end{equation}
and $T_{ij}^\text{C}(t)$ is the time required to execute the task at the UAV as
\begin{equation}\label{w11}
\begin{aligned}
T_{ij}^\text{C}(t) = \frac{F_i(t)}{f^\text{C}_{ij}(t)},~\forall t \in \mathcal{T},
\end{aligned}
\end{equation}
where $f_{ij}^{\text{C}}(t)$ is the computation resource that the $j$-th UAV can provide to the $i$-th UE in the $t$-th time slot.

Note that the time needed for returning the results back to UE from UAV is ignored, similar to~\cite{8353131}. The overall energy consumption of the $i$-th UE to the $j$-th UAV in the $t$-th time slot is given by
\begin{equation}\label{w12}
\begin{aligned}
E^{\text{Tr}}_{ij}(t) = P^{\text{Tr}}T_{ij}^{\text{Tr}}(t),~\forall t \in \mathcal{T}.
\end{aligned}
\end{equation}
\textcolor{black}{If the UE decides to execute the task locally, the power consumption can be evaluated as $k_i{(f_{ij}^{\text{L}}(t))}^{v_i}$~\cite{jiang2019deep}, where $k_i \geq 0$ is the effective switched capacitance, $v_i$ is typically set to 3, and $f_{ij}^{\text{L}}(t)$ is the computation resource that the $i$-th UE applies to execute the task. The overall time for local execution can be given by}
\begin{equation}\label{we12}
\begin{aligned}
T_{ij}^\text{L}(t) = \frac{F_i(t)}{f^\text{L}_{ij}(t)}.
\end{aligned}
\end{equation}

Thus, the total energy consumption for local execution is 
\begin{equation}\label{2}
\begin{aligned}
E_{ij}^\text{L}(t) =k_i(f^\text{L}_{ij}(t))^{v_i}T_{ij}^\text{L}(t),~ t \in \mathcal{T}.
\end{aligned}
\end{equation}
 
To sum up, the overall energy consumption for task execution $E_{ij}(t)$ is given by
\begin{equation}
\begin{aligned}[l]
E_{ij}(t) = 
\begin{cases}
E_{ij}^\text{L}(t), & \text{local execution}, \\
E_{ij}^{\text{Tr}}(t), & \text{offloading}, \\
\end{cases}
\end{aligned}
\end{equation}
and the time to complete the task $T_{ij}(t)$ is expressed as
\begin{equation}
\begin{aligned}[l]
T_{ij}(t) = 
\begin{cases}
T_{ij}^\text{L}(t), & \text{local execution}, \\
T_{ij}^\text{O}(t), & \text{offloading}. \\
\end{cases}
\end{aligned}
\end{equation}

Without loss of generality, we assume that each task has to be completed within  maximal time duration $T^{\text{max}}$, which is consistent with the maximal flying time in each time slot as
\textcolor{black}{\begin{equation}\label{w13}
		\begin{aligned}
			T_{ij}(t) \leq T^{\text{max}},~ \forall i \in \mathcal{N}, j \in \mathcal{M}', t \in \mathcal{T}.
		\end{aligned}
\end{equation}}

\textcolor{black}{In each time slot, since the computation resource that each UAV can provide is limited, we have}
\textcolor{black}{\begin{equation}\label{w14}
		\begin{aligned}
			\sum_{i=1}^{N}a_{ij}(t)f_{ij}^{\text{C}}(t) \leq f^{\text{max}},~ \forall j \in \mathcal{M}, t \in \mathcal{T},
		\end{aligned}
\end{equation}}
\textcolor{black}{where $f^{\text{max}}$ is the maximal computation resource that the $j$-th UAV can provide in each time slot. Next, we show our proposed problem formulation.}

\subsection{Problem Formulation}
Denote $\bm{U}$ = $\{\theta^h_j(t), d_j(t), \forall j \in \mathcal{M}, t \in \mathcal{T}\}$, $\bm{A}$ = $\{a_{ij}(t), \forall i \in \mathcal{N}, j \in \mathcal{M}', t \in \mathcal{T}\}$, $\bm{F}$ = $\{f_{ij}(t), \forall i \in \mathcal{N}, j \in \mathcal{M}', t \in \mathcal{T}\}$. Then, the energy minimization for all UEs is formulated as

\textcolor{black}{\begin{subequations}\label{ww14}
		\begin{IEEEeqnarray}{s,lCl'lCl'lCl}
			& \IEEEeqnarraymulticol{9}{l}{\mathcal{P}1: \underset{\bm{U},\bm{A},\bm{F}}{\text{min}}~ 
				\sum_{i=1}^{N}\sum_{j=0}^{M}\sum_{t=1}^{T}a_{ij}(t)E_{ij}(t)} \\
			& \text{subject to:} \nonumber\\
			& a_{ij}(t) = \{0, 1\}, \forall i \in \mathcal{N}, j \in \mathcal{M}', t \in \mathcal{T}, \\
			&\sum_{j=0}^{M}a_{ij}(t)=1, \forall i \in \mathcal{N}, t \in \mathcal{T}, \\
			&\sum_{i=1}^{N}a_{ij}(t) \leq V^{\text{max}}, \forall j \in \mathcal{M}, t \in \mathcal{T},\\
			&0 \leq \theta^h_{j}(t) \leq 2\pi , \forall j \in \mathcal{M}, t \in \mathcal{T},\\
			&0 \leq d_{j}(t) \leq d^{\text{max}}, \forall j \in \mathcal{M}, t \in \mathcal{T},\\
			&0 \leq X_j(t) \leq X^{\text{max}}, \forall j \in \mathcal{M}, t \in \mathcal{T},\\
			&0 \leq Y_j(t) \leq Y^{\text{max}}, \forall j \in \mathcal{M}, t \in \mathcal{T},\\
			&a_{ij}(t)R_{ij}(t) \leq R^{\text{max}}, \forall i \in \mathcal{N}, j \in \mathcal{M}, t \in \mathcal{T},\\
			& T_{ij}(t) \leq T^{\text{max}},~ \forall i \in \mathcal{N}, j \in \mathcal{M}', t \in \mathcal{T},\\
			& \sum_{i=1}^{N}a_{ij}(t)f_{ij}^{\text{C}}(t) \leq f^{\text{max}},~ \forall j \in \mathcal{M}, t \in \mathcal{T}.
		\end{IEEEeqnarray}
\end{subequations}}

\textcolor{black}{One can see that the above problem $\mathcal{P}1$ is a mixed integer
	nonlinear programming (MINLP), as it includes both integer variable, $\bm{A}$ and continuous variables, $\bm{F}$ and $\bm{U}$, which is very difficult to solve in general. We first propose a convex optimization based algorithm CAT to address it iteratively. Then, we propose a Deep Reinforcement Learning (DRL) based RAT to facilitate fast decision-making, which can be applied in dynamic environment.
	Note that in practice, if the $i$-th UE does not generate the tasks in the $t$-th time slot and then the corresponding $D_i(t)$ and $F_i(t)$ can be set to zero.}

\section{Proposed CAT Algorithm}

In this section, a convex optimization based CAT is proposed to solve the above problem $\mathcal{P}1$. We first define a set of new variables to denote the trajectories of UAVs as $\bm{G}$ = $\{G_j(t), \forall j \in \mathcal{M}, t \in \mathcal{T}\}$, where the coordinate is $G_j(t) = [X_j(t), Y_j(t)]$, $X_{j}(t)=X_{j}(0)+\sum_{l=1}^{t}d_{j}(l)\text{cos}\big(\theta^h_{j}(l)\big)$ and $Y_{j}(t)=Y_{j}(0)+\sum_{l=1}^{t}d_{j}(l)\text{sin}\big(\theta^h_{j}(l)\big)$. Thus, the optimization problem $\mathcal{P}1$ can be reformulated as

\textcolor{black}{\begin{subequations}\label{cvx}
		\begin{IEEEeqnarray}{s,lCl'lCl'lCl}
			& \IEEEeqnarraymulticol{9}{l}{\mathcal{P}2: \underset{\bm{G},\bm{A},\bm{F}}{\text{min}}~ 
				\sum_{i=1}^{N}\sum_{j=0}^{M}\sum_{t=1}^{T}a_{ij}(t)E_{ij}(t)} \\
			& \text{subject to:}~ (\ref{ww14}\text{b}), (\ref{ww14}\text{c}), (\ref{ww14}\text{d}), (\ref{ww14}\text{g}), (\ref{ww14}\text{h}), (\ref{ww14}\text{j}), (\ref{ww14}\text{k}),\nonumber\\
			& a_{ij}(t)||G_j(t)-q_i||^2 \leq (R^{\text{max}})^2, \forall i \in \mathcal{N}, j \in \mathcal{M}, t \in \mathcal{T},\\
			& ||G_j(t+1)-G_j(t)||^2 \leq (d^{\text{max}})^2, \forall t \in \{0,1,..., T-1\},
		\end{IEEEeqnarray}
\end{subequations}}
\textcolor{black}{where $q_i= [x_i,y_i]$. In order to solve $\mathcal{P}2$, we divide it into two subproblems and apply the block coordinate descent (BCD) method to address it. To this end, we first optimize the user association $\bm{A}$ and resource allocation $\bm{F}$ given the UAV trajectory $\bm{G}$. Then, we optimize the UAV trajectory $\bm{G}$ given the user association $\bm{A}$ and resource allocation $\bm{F}$. We solve the two optimization problems iteratively, until the convergence is achieved.}

\subsection{User Association and Resource Allocation}
Given the UAV trajectory $\bm{G}$, the subproblem to decide user association $\bm{A}$ and resource allocation $\bm{F}$ can be formulated as
\textcolor{black}{\begin{subequations}\label{AF}
		\begin{IEEEeqnarray}{s,lCl'lCl'lCl}
			& \IEEEeqnarraymulticol{9}{l}{\underset{\bm{A},\bm{F}}{\text{min}}~ 
				\sum_{i=1}^{N}\sum_{j=0}^{M}\sum_{t=1}^{T}a_{ij}(t)E_{ij}(t)} \\
			& \text{subject to:}~(\ref{ww14}\text{b}), (\ref{ww14}\text{c}), (\ref{ww14}\text{d}), (\ref{ww14}\text{j}), (\ref{ww14}\text{k}),(\ref{cvx}\text{b}). \nonumber
		\end{IEEEeqnarray}
\end{subequations}}
One can see that (\ref{ww14}j) can be written as
\begin{equation}\label{w16}
\begin{aligned}
f_{ij}^{\text{C}}(t) \geq \frac{F_i(t)}{T^{\text{max}}-\frac{D_i(t)}{r_{ij}(t)}},~ \forall j \in \mathcal{M}, t \in \mathcal{T},
\end{aligned}
\end{equation}
if the $i$-th UE chooses to offload the task, and
\begin{equation}\label{w17}
\begin{aligned}
f_{ij}^{\text{L}}(t) \geq \frac{F_i(t)}{T^{\text{max}}},~ j = 0 , \forall t \in \mathcal{T},
\end{aligned}
\end{equation}
if the $i$-th UE decides to execute the task locally. It is readily to see that equality holds for both (\ref{w16}) and (\ref{w17}).

Then, (\ref{AF}) can be re-written as 
\textcolor{black}{\begin{subequations}\label{AF1}
		\begin{IEEEeqnarray}{s,lCl'lCl'lCl}
			& \IEEEeqnarraymulticol{9}{l}{\underset{\bm{A},\bm{F}}{\text{min}}~ 
				\sum_{i=1}^{N}\sum_{j=0}^{M}\sum_{t=1}^{T} \left( a_{ij}(t)E_{ij}^\text{Tr}(t)+ ( 1-a_{ij}(t))E_{ij}^\text{L}(t)\right)} \\
			& \text{subject to:}~(\ref{ww14}\text{b}), (\ref{ww14}\text{c}), (\ref{ww14}\text{d}), (\ref{cvx}\text{b}), \nonumber\\
			& f_{ij}^{\text{L}}(t) = \frac{F_i(t)}{T^{\text{max}}},~ j = 0 , \forall t \in \mathcal{T},\\
			& \sum_{i=1}^{N}a_{ij}(t) \frac{F_i(t)}{T^{\text{max}}-\frac{D_i(t)}{r_{ij}(t)}} \leq f^{\text{max}},~ \forall j \in \mathcal{M}, t \in \mathcal{T}.
		\end{IEEEeqnarray}
\end{subequations}}
It is ready to find (\ref{AF1}) is similar to a Multiple-Choice Multi-Dimensional 0-1 Knapsack Problem (MMKP), which is difficult to solve in general. Fortunately, it may be addressed by applying Branch and Bound method via a standard Python package PULP~\cite{Mitchell2011PuLPA}.  

\subsection{UAV Trajectory Optimization}
Given the user association and resource allocation from (\ref{AF1}) and removing the constant, $\mathcal{P}2$ can be simplified as
\textcolor{black}{\begin{subequations}\label{G}
		\begin{IEEEeqnarray}{s,lCl'lCl'lCl}
			& \IEEEeqnarraymulticol{9}{l}{\underset{\bm{G}}{\text{min}}~ 
				\sum_{i=1}^{N}\sum_{j=1}^{M}\sum_{t=1}^{T} a_{ij}(t) \frac{P^{\text{Tr}} D_i(t)}{B \text{log}_2(1+\frac{\alpha P^{\text{Tr}}}{Z_{j}^2+||G_j(t)-q_i||^2})}}\\
			& \text{subject to:}~(\ref{ww14}\text{g}), (\ref{ww14}\text{h}), (\ref{cvx}\text{b}), (\ref{cvx}\text{c}), \nonumber\\
			& \frac{D_i(t)}{B \text{log}_2(1+\frac{\alpha P^{\text{Tr}}}{Z_{j}^2+||G_j(t)-q_i||^2})}+\frac{F_i(t)}{f^{\text{C}}_{ij}(t)} \leq T^{\text{max}}, \nonumber\\&\quad\quad\quad\quad\quad\quad\quad\quad\quad\quad \forall i \in \mathcal{N}, j \in \mathcal{M}, t \in \mathcal{T}.
		\end{IEEEeqnarray}
\end{subequations}}
It is easy to see that the above optimization problem is non-convex with respect to $G_j(t)$. Next, we introduce a set $\bm{\eta} = \{\eta_{ij}(t) , \forall i \in \mathcal{N}, j \in \mathcal{M}, t \in \mathcal{T}\}$, where $\eta_{ij}(t) =  a_{ij}(t) \frac{P^{\text{Tr}} D_i(t)}{B \text{log}_2(1+\frac{\alpha P^{\text{Tr}}}{Z_{j}^2+||G_j(t)-q_i||^2})}$, then, problem (\ref{G}) can be transformed into

\textcolor{black}{\begin{subequations}\label{cvxG}
		\begin{IEEEeqnarray}{s,lCl'lCl'lCl}
			& \IEEEeqnarraymulticol{9}{l}{\underset{\bm{G},\bm{\eta}}{\text{min}}~ 
				\sum_{i=1}^{N}\sum_{j=1}^{M}\sum_{t=1}^{T} \eta_{ij}(t)}\\
			& \text{subject to:}~(\ref{ww14}\text{g}), (\ref{ww14}\text{h}), (\ref{cvx}\text{b}), (\ref{cvx}\text{c}), \nonumber\\
			&B \text{log}_2\big(1+\frac{\alpha P^{\text{Tr}}}{Z_{j}^2+||G_j(t)-q_i||^2}\big) \geq \frac{a_{ij}(t)P^{\text{Tr}} D_i(t)}{\eta_{ij}(t)}, \nonumber\\& \quad\quad\quad\quad\quad\quad\quad\quad\quad\quad \forall i \in \mathcal{N}, j \in \mathcal{M}, t \in \mathcal{T},\\
			& B \text{log}_2\big(1+\frac{\alpha P^{\text{Tr}}}{Z_{j}^2+||G_j(t)-q_i||^2}\big) \geq \frac{D_i(t)}{T^{\text{max}}-\frac{F_i(t)}{f_{ij}^{\text{C}}(t)}}, \nonumber\\&\quad\quad\quad\quad\quad\quad\quad\quad\quad\quad \forall i \in \mathcal{N}, j \in \mathcal{M}, t \in \mathcal{T}.
		\end{IEEEeqnarray}
\end{subequations}}

One observes that (\ref{cvxG}b) and (\ref{cvxG}c) are convex with respect to $||G_j(t)-q_i||$, respectively. Thus, (\ref{cvxG}b) and (\ref{cvxG}c) are non-convex constraints. 
Then, similar to~\cite{8438896,8637952}, we apply the successive
convex approximation (SCA) to solve this problem. Specifically,  for any given local point $G_j^{r}(t)$ in $\bm{G}^r = \{G_j^r(t), \forall j \in \mathcal{M}, t \in \mathcal{T}\}$, one can have the following inequality as
\begin{equation}
\begin{aligned}
w_{ij}(t) & = B \text{log}_2\big(1+\frac{\alpha P^{\text{Tr}}}{Z_{j}^2+||G_j(t)-q_i||^2}\big) \\&\geq K^{r}_{ij}(t)( ||G_j(t)-q_i||^2 - ||G^{r}_j(t) - g_i||^2 ) + B^{r}_{ij}(t) \\&\triangleq w^{lb,r}_{ij}(t), 
\end{aligned}
\end{equation}
where
$
K^{r}_{ij}(t) = -\frac{B \alpha P^{\text{Tr}} \text{log}_2(e)}{(Z_j^2 + ||G^{r}_j(t)-q_i||^2) (Z_j^2 + ||G^{r}_j(t)-q_i||^2 + \alpha P^{\text{Tr}})},
$
and
$
B^{r}_{ij}(t) = B\text{log}_2\big(1+ \frac{\alpha P^{\text{Tr}}}{Z_j^2 +||G^{r}_j(t)- q_i||^2}\big)
$.

Then, problem (\ref{cvxG}) can be written as
\textcolor{black}{\begin{subequations}\label{Gcon}
		\begin{IEEEeqnarray}{s,lCl'lCl'lCl}
			& \IEEEeqnarraymulticol{9}{l}{\underset{\bm{G},\bm{\eta}}{\text{min}}~ 
				\sum_{i=1}^{N}\sum_{j=1}^{M}\sum_{t=1}^{T}\eta_{ij}(t)}\\
			& \text{subject to:}~ (\ref{ww14}\text{g}), (\ref{ww14}\text{h}), (\ref{cvx}\text{b}), (\ref{cvx}\text{c}), \nonumber\\
			&w^{lb,r}_{ij}(t) \geq \frac{a_{ij}(t)P^{\text{Tr}} D_i(t)}{\eta_{ij}(t)} , \forall i \in \mathcal{N}, j \in \mathcal{M}, t \in \mathcal{T},  \\
			& w^{lb,r}_{ij}(t) \geq \frac{D_i(t)}{T^{\text{max}}-\frac{F_i(t)}{f_{ij}^{\text{C}}(t)}}, \forall i \in \mathcal{N}, j \in \mathcal{M}, t \in \mathcal{T}.
		\end{IEEEeqnarray}
\end{subequations}}

The above problem is a convex quadratically constrained quadratic program (QCQP) and it can be solved by a standard Python package CVXPY~\cite{cvxpy}.

\subsection{Overall Algorithm Design}
In this section, a convex optimization-based CAT is proposed to solve Problem $\mathcal{P}2$, where we optimize user association and resource allocation subproblem iteratively with the UAV trajectory subproblem until the convergence is achieved. We describe the pseudo code of proposed CAT in Algorithm \ref{CVX}.

\begin{algorithm}[!htpb]
	\caption{CAT Algorithm}\label{CVX}
	\begin{algorithmic}[1] 
		\STATE Set $r=0$, and initialize $\bm{G}^{r}$; \\
		\STATE \textbf{repeat} \
		\STATE Solve Problem (\ref{AF1}) by Branch and Bound method for given $\bm{G}^{r}$, and denote the optimal solution as $\bm{A}^{r+1}$ and $\bm{F}^{r+1}$;\
		\STATE Solve Problem (\ref{Gcon}) for given $\bm{A}^{r+1}$ and $\bm{F}^{r+1}$, and denote the solution as $\bm{G}^{r+1}$;
		\STATE $r = r + 1$;\
		\STATE \textbf{until} the convergence is achieved. \
	\end{algorithmic} 
\end{algorithm}

{\bf{Discussions:}}  Algorithm 1 needs to run once the initial taking off locations of the  UAVs change.  However, the complexity of Algorithm 1 is high as the solutions are iteratively obtained and each subproblem involves a huge number of optimization variables especially when the total number of time slots is high. Precisely, as shown in Algorithm~\ref{CVX}, assume that the overall iteration number is $K^r$. In each iteration, Problem (\ref{AF1}) has $N(M+1)T$ variables, and it can be solved by Branch and Bound method, in which the Simplex technique for solving linear programs is used. Thus, the computational complexity is $\mathcal{O}\big(2^{N(M+1)T}\big)$ in the worst case. \textcolor{black}{Furthermore, according to the analysis in~\cite{8438896, pan2017joint}, in Problem (\ref{Gcon}), $\bm{G}$ has $2MT$ variables, $\bm{\eta}$ has $NMT$ variables. Hence, the total number of variables is $(N+2)MT$. As a result, the number of iterations required is $\mathcal{O}\big(\sqrt{(N+2)MT}\text{log}_2(\frac{1}{\epsilon_1})\big)$, where $\epsilon_1$ is the accuracy of SCA for solving Problem (\ref{Gcon}). Similarly, the overall number of constraints in Problem (\ref{Gcon}) is $MT(3N+2)+T$. Then, the computational complexity is $\mathcal{O} \bigg(\big((N+2)MT\big)^2\sqrt{(N+2)MT}  \text{log}_2(\frac{1}{\epsilon_1})\big(MT(3N+2)+T\big)\bigg)$, which is equivalent to  $\mathcal{O}\big(3(NMT)^{3.5}\text{log}_2(\frac{1}{\epsilon_1})\big)$. Overall, the total complexity of CAT algorithm is $\mathcal{O}\big(K^r\big(2^{N(M+1)T}+3(NMT)^{3.5}\text{log}_2(\frac{1}{\epsilon_1})\big)\big)$.} Hence, Algorithm 1 is not suitable for some emergence scenarios (e.g., battlefields, earthquake, large fires), where fast decision making is highly demanded. This motivates the  algorithm developed based on DRL in  the following section.

\section{Proposed RAT Algorithm}
To facilitate the fast decision making, the DRL-based RAT algorithm is proposed in this section.
We first give some preliminaries as follows.

\subsection{Preliminaries}

\subsubsection{DQN}
In a standard reinforcement learning, an agent is assumed to interact with the environment and select the optimal actions that can maximize the accumulated reward. 
In~\cite{mnih2015human}, a Deep Q Network (DQN) structure developed by Google Deepmind, integrates the deep neural networks with traditional reinforcement learning. The DQN is used to estimate the well-known Q-value defined as
\begin{equation}\label{w18}
\begin{aligned}
Q(s(t), c(t)) = \mathbb{E}[Z(t)|s(t), c(t)],
\end{aligned}
\end{equation}
where $s(t)$ and $c(t)$ denote the state and action respectively, $\mathbb{E}[\cdot]$ denotes the expectation, whereas $Z(t) = \sum_{t'=t}^{T} \gamma z(t')$ is a reward and $\gamma \in [0, 1]$ is the discount factor and $z(t')$ is a reward function in the $t'$-th time step (or time slot). As the objective is to maximize the reward, a widely used policy is $\pi(s(t)|\phi^Q) =\text{ argmax}_{c(t)}Q(s(t), c(t)|\phi^Q)$, where $\phi^Q$ is the parameter of the deep neural network. Then, the DQN can be trained by minimizing the loss function~\cite{mnih2015human}. Also, since the deep networks are known to be unstable and very difficult to converge, two effective approaches, i.e., target network and experience replay, have been introduced in \cite{mnih2015human}. The target network has the same structure as the original DQN but the parameters are updated more slowly. The experience replay stores the state transition samples which can help the DQN converge. However, the DQN was originally designed to solve the problem with discrete variables. Although we can adapt the DQN to continuous problems by discretizing the action space, it may unfortunately result in a huge searching space and therefore intractable to deal with.         

\subsubsection{DDPG}
To deal with the problem with continuous variables, e.g., the trajectory control of UAV, one may apply the actor-critic approach, which was developed in~\cite{konda2000actor}. DeepMind has proposed a deep deterministic policy gradient (DDPG) approach~\cite{lillicrap2015continuous} by integrating the actor-critic approach into DRL. DDPG includes two DQNs, one of the DQNs, named actor network with function $\pi(s(t)|\phi^{\pi})$ is applied to generate action $c(t)$ for a given state $s(t)$. The other DQN named critic network with function $Q(s(t), c(t)|\phi^Q)$, is used to generate the Q-value, which evaluates the action produced by the actor network. In order to improve the learning stability, two adjacent target networks corresponding to the actor and critic networks, $\pi'(\cdot)$, $Q'(\cdot)$ with respective parameters $\phi^{\pi'}$, $\phi^{Q'}$, are also applied. 

Then, the critic network can be updated with the loss function, $L(\phi^Q)$, as
\begin{equation}\label{w19}
\begin{aligned}
L(\phi^Q) = \frac{1}{K}\sum_{k=1}^{K}\delta_k^2,
\end{aligned}
\end{equation}
where in each time step, the mini-batch randomly samples $K$ constituting experiences from experience replay buffer, and $\delta_k$ is temporal difference (TD)-error~\cite{van2016deep} which is given by
\begin{equation}\label{w20}
\begin{aligned}
\delta_k = &z(k) + \gamma Q'(s(k+1), \pi'(s(k+1)|\phi^{\pi'})|\phi^{Q'}) \\&- Q(s(k), \pi(s(k)|\phi^{\pi})|\phi^Q).
\end{aligned}
\end{equation} 

On the other hand, the actor network can be updated by applying the policy gradient, which is described as~\cite{lillicrap2015continuous}.
\begin{equation}\label{w21}
\begin{aligned}
\bigtriangledown_{\phi^{\pi}}J & \approx \frac{1}{K}\sum_{k=1}^{K}\bigtriangledown_{c}Q(s, c|\phi^{Q})|_{s=s(k), c=\pi(s(k)|\phi^{\pi})} = \\& \frac{1}{K}\sum_{k=1}^{K}\left[\bigtriangledown_{c}Q(s, c|\phi^Q)|_{s=s(k), c=\pi(s(k))} \cdot \bigtriangledown_{\phi^{\pi}}\pi(s|\phi^{\pi})|_{s=s(k)}\right].\\
\end{aligned}
\end{equation}

\subsection{The RAT Algorithm}\label{section3}

In this section, we introduce the DRL based RAT algorithm, which includes deep neural networks (i.e., actor and critic networks) and the matching algorithms. In order to apply the DRL, we first define the state, action and reward as follows:
\begin{itemize}
	\item[1)] \textcolor{black}{\textbf{State} $s(t)$: $s(t) = \{[X_j(t), Y_j(t), Z_j],~\forall j \in \mathcal{M}\}$, $s(t)$ is the set of the coordinates of all UAVs.}
	
	\item[2)] \textcolor{black}{\textbf{Action} $c(t)$: $c(t)$ is the set of the actions of all UAVs, including the horizontal direction $\theta^h_j(t)$ and distance $d_j(t)$. Then, the action set can be defined as $c(t) = \{[\theta^h_j(t), d_j(t)],~\forall j \in \mathcal{M}\}$.}
	
	\item[3)] \textcolor{black}{\textbf{Reward} $z(t)$: $z(t)$ is defined as the minus of the overall energy consumption of all the UEs in each time slot as
		\begin{equation}\label{ww18}
			z(t) = - \sum_{i=1}^{N}\sum_{j=0}^{M}a_{ij}(t)E_{ij}(t) - p,
	\end{equation}}
     \textcolor{black}{where $p$ is the penalty if any of UAV flies out of the target area, which means (\ref{ww14}g) or (\ref{ww14}h) is not satisfied.}
\end{itemize}

\begin{figure}[htpb]
	\centering
	\includegraphics[width=3.6in]{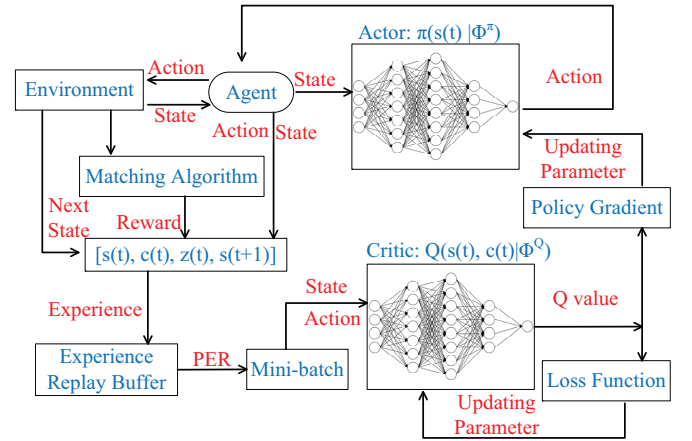}
	\caption{The structure of RAT algorithm.}\label{dia}
\end{figure}

The algorithm framework used in this paper is depicted in Fig. \ref{dia}, where an agent, which could be deployed in the control center of the base station, is assumed to interact with the environment. An actor network $\pi(s(t)|\phi^{\pi})$ is applied to generate the action, which includes the flying direction and distance for each UAV. The critic network $Q(s(t), c(t)|\phi^Q)$ is used to obtain the Q-value of the action (i.e., to evaluate the action generated by actor network). In each time slot, the agent sends the action generated by actor network to each UAV. Then, each UE tries to associate with one UAV in its coverage, i.e., (\ref{ww8}) by using a matching algorithm in Algorithm~\ref{match}. More specifically, each UE tries to connect the UAV which can save more offloading energy. If the minimum offloading energy is larger than the energy of local execution, the UE will decide to conduct the task locally. Note that RAT has the same optimization strategy for resource allocation as CAT.

\textcolor{black}{Also, each UAV selects the UEs based on the following criteria: 1) UE should be within its coverage area; 2) UE could save more energy, i.e., the more of $E^{\text{L}}_{ij}(t)-E^{\text{C}}_{ij}(t)$ will be given higher priority in offloading to this UAV. We will introduce the details of the proposed matching algorithm in Algorithm~\ref{match}. After the matching algorithm, the reward in (\ref{ww18}) can be obtained.  }

We assume that there is an experience replay buffer for the agent to store the experience $[s(t), c(t), z(t), s(t+1)]$. Once the experience replay buffer is full, the learning procedure starts. A mini-batch with $K$ experiences can be obtained from the experience replay buffer to train the networks. 

In the classical DRL algorithms, such as Q-learning~\cite{watkins1992q}, SARSA~\cite{hamari2014does} and DDPG~\cite{lillicrap2015continuous}, the mini-batch uniformly samples experiences from the experience replay buffer. However, since TD-error in (\ref{w20}) is used to update the Q-value network, experience with high TD-error often indicates the successful attempts. Therefore, a better way to select the experience is to assign different weights to samples. Schaul~\emph{et al.}~\cite{schaul2015prioritized} developed a prioritized experience replay scheme, in which the absolute TD-error $|\delta_k|$ is used to evaluate the probability of the sampled $k$-th experience from the mini-batch. Then, the probability of sampling the $k$-th experience can be given by
\begin{equation}\label{w22}
\begin{aligned}
P(k) = \frac{p_k^{\beta}}{\sum_{m \in K}p_m^{\beta}},
\end{aligned}
\end{equation}
where $p_k = |\delta_k|+\epsilon$, $\epsilon = 0.001$ is a positive constant to avoid the edge-case of transitions not being revisited if $|\delta_k|$ is 0, $\beta = 0.6$ is denoted as a factor to determine the prioritization~\cite{schaul2015prioritized}.

However, frequently sampling experiences with high $|\delta_k|$ can cause divergence and oscillation. To tackle this issue, the importance-sampling weight~\cite{mahmood2014weighted} is introduced to represent the importance of sampled experience, which can be given by 

\begin{equation}\label{w23}
\begin{aligned}
\omega_k = \frac{1}{(X\cdot P(k))^{\mu}},
\end{aligned}
\end{equation}
where $X$ is size of experience replay buffer, $\mu$ is given as 0.4~\cite{schaul2015prioritized}. Thus, the loss function $L(\phi^Q)$ in (\ref{w19}) is updated as
\begin{equation}\label{w24}
\begin{aligned}
L(\phi^Q) = \frac{1}{K}\sum_{k=1}^{K}\omega_k\delta_k^2,
\end{aligned}
\end{equation}
which is used in our proposed RAT to train the networks. Next, we describe the pseudo code of the overall RAT framework in Algorithm~\ref{RAT}.

\begin{algorithm}[!htpb]
	\caption{RAT Algorithm}\label{RAT}
	\begin{algorithmic}[1] 
		\STATE Initialize actor network $\pi(s(t)|\phi^{\pi})$ with parameters $\phi^{\pi}$ and critic network $Q(s(t), s(t)|\phi^Q)$ with parameters $\phi^Q$; \
		\STATE Initialize target networks $Q'(\cdot)$ with parameters $\phi^{Q'} = \phi^{Q}$ and $\pi'(\cdot)$ with parameters $\phi^{\pi'} = \phi^{\pi}$; \
		\STATE Initialize experience replay buffer $\mathcal{X}$; \
		\FOR{epoch =1,..., $k^{\text{max}}$}
		\STATE Initialize $s(t)$; \
		\FOR{time slot $t$ =1,..., $T$}
		\STATE $\pi(s(t)|\phi^{\pi}) + \rho N'$ where $N'$ is the random noise and $\rho$ decays with $t$; \
		\FOR{UAV $j$=1,..., $M$}
		\STATE Execute $c(t)$; \
		\STATE Obtain $s(t+1)$; \
		\ENDFOR
		\STATE Obtain the user association with UAVs using matching algorithm proposed in Algorithm~\ref{match}; \
		\STATE Obtain the reward $z(t)$ from (\ref{ww18}); \
		\STATE Store experience [$s(t), c(t), z(t), s(t+1)$] into the replay buffer; \
		\IF{the replay buffer is full}
		\FOR{$k$ = 1,..., $K$}
		\STATE Sample $k$-th experience with probability $P(k)$ from (\ref{w22}); \
		\STATE Calculate $|\delta_k|$ and $\omega_k$ from (\ref{w20}) and (\ref{w23}) respectively; \
		\ENDFOR
		\STATE Update parameters of the critic network $\phi^{Q}$ by minimizing its loss function according to (\ref{w24}); \
		\STATE Update parameters of the actor network $\phi^{\pi}$ by using policy gradient approach according to (\ref{w21}); \
		\STATE Update two target networks with the updating rate $\tau$: \
		\ENDIF
		\ENDFOR
		\ENDFOR
	\end{algorithmic} 
\end{algorithm}

We first initialize the actor, critic, two target networks, and experience replay buffer in Line 1 - 3. In the beginning of each epoch, all UAVs start to serve UEs from different taking off points. Note that for better exploration, we add a random noise $N'$ to the action, where $N'$ follows a normal distribution with $0$ mean and variance $1$, $\rho$ is set to 2 and decays with a rate of 0.9995 in each time step. From Line 8-11, each UAV flies according to the generated action $c(t)$ and enters the next state $s(t+1)$. Then, we obtain the user association by using Algorithm~\ref{match}. Next, the reward $z(t)$ is obtained according to (\ref{ww18}) (i.e., Line 13). The experience is also stored in the replay buffer. When the buffer is full, the mini-batch samples $K$ experiences by applying the prioritized experience replay (i.e., Line 16-19). Then, we update the actor and critic networks by using loss function in (\ref{w24}) and policy gradient in (\ref{w21}) respectively. Finally, we update the target networks by using the following equations as (i.e., Line 22)

\begin{equation}\label{ww23}
\begin{aligned}
\phi^{Q'} \leftarrow \tau\phi^{Q} + (1-\tau)\phi^{Q'},
\end{aligned}
\end{equation} 
and
\begin{equation}\label{we3}
\begin{aligned}
\phi^{\pi'} \leftarrow \tau\phi^{\pi} + (1-\tau)\phi^{\pi'},
\end{aligned}
\end{equation}  
where $\tau$ is the updating rate.

\begin{algorithm}[!htpb]
	\caption{Matching Algorithm}\label{match}
	\begin{algorithmic}[1]
		\STATE Initialize $\bm{A}$ and $\bm{F}_j$, $\forall j \in \mathcal{M}$, $\forall i \in \mathcal{N}$; \
		\FOR{UAV $j$ = 1,..., $M$}
		\FOR{UE $i$ = 1,..., $N$}
		\IF{(\ref{ww8}) is met}
		\STATE Calculate $E_{ij}^{\text{L}}(t)$, $E_{ij}^{\text{Tr}}(t)$ and $f_{ij}^{\text{C}}(t)$; \
		\IF{$E_{ij}^{\text{L}}(t) > E_{ij}^{\text{Tr}}(t)$}
		\STATE Store $i$ into $\bm{E}_j$;
		\ENDIF
		\ENDIF
		\ENDFOR
		\STATE Sort the element in $\bm{E}_j$ in descending order with respect to $E_{ij}^{\text{L}}(t)-E_{ij}^{\text{Tr}}(t)$; \
		\ENDFOR
		\STATE \textbf{repeat} \
		\FOR{UAV $j$ = 1,..., $M$} 
		\STATE $i = GetTopItem(\bm{E}_j)$; \
		\IF{(\ref{w6}), (\ref{w14}) are met}
		\IF{$E_{ij}^{\text{Tr}}(t) < E_{i\bm{A}(i)}^{\text{Tr}}(t)$ or $\bm{A}(i) = 0$}
		\STATE $\bm{A}(i) = j$; \
		\ENDIF
		\STATE $RemoveTopItem(\bm{E}_j)$; \
		\ENDIF
		\ENDFOR
		\STATE \textbf{until} Each UE in $\bm{E}_j$ is checked. \
		\STATE Return $\bm{A}$\
	\end{algorithmic}
\end{algorithm}

\textcolor{black}{Next, we introduce the low-complexity matching algorithm which can decide the user association and resource allocation given UAVs' trajectories, as shown in Algorithm~\ref{match}. First, we denote $\bm{A}$ with size $N$ to record the user association between UEs and UAVs. If $\bm{A}(i) = j$, the $i$-th UE matches with the $j$-th UAV, while if $\bm{A}(i) = 0$, the $i$-th UE is not matched yet and has to execute its task locally. In addition, we denote a preference list $\bm{E}_j$ for the $j$-th UAV to record UEs that can benefit from offloading. Then, from Line 2 to 10, we generate the preference list $\bm{E}_j$ for the $j$-th UAV. Precisely, if constraint (\ref{ww8}) is met, we obtain $E_{ij}^{\text{L}}(t)$, $E_{ij}^{\text{Tr}}(t)$, and $f_{ij}^{\text{C}}(t)$ according to (\ref{2}), (\ref{w12}), and (\ref{w16}), respectively. UEs that benefit from offloading will be stored in $\bm{E}_j$. Since UAVs need to save as much energy of UEs as possible, we sort the preference list $\bm{E}_j$ with descending order with respect to $E_{ij}^{\text{L}}(t)-E_{ij}^{\text{Tr}}(t)$, as shown in Line 11. The UE that can save more energy via offloading will be matched with a higher priority. Next, from Line 13 to 23, we conduct the matching process. Each UAV keeps selecting UEs according to its preference list, and constantly checking the constraints (\ref{w6}) and (\ref{w14}) based on $\bm{A}$. In the meantime, the selected UE will determine whether to match with the UAV or not. Precisely, from Line 17 to 19, if the selected UE is not matched before, or matching with the $j$-th UAV could save more energy than previous match, the corresponding $\bm{A}(i)$ will be updated. We do this process until all the UEs in each preference list are checked. Then, the final user association can be obtained from $\bm{A}$.}

\textcolor{black}{According to~\cite{lillicrap2015continuous}, our RAT algorithm is an offline learning and off-policy DRL-based algorithm as the experience replay mechanism is applied, and the mini-batch will sample several uncorrelated experiences for training networks in each time step. Additionally, the training procedure can be deployed in a simulator, and the RAT model can be easily deployed in reality when the convergence is achieved, which will inevitably reduce the payoff of implementation.} Furthermore, once the whole networks are converged, the solutions can be generated very fast with only some simple algebraic calculations instead of solving the original MINLP. This is due to the fact that during the training stages, random taking off points of all the UAVs are generated and the networks are trained to converge.   

\textcolor{black}{{\bf{Discussions:}} after adequate training process, the RAT model, including the networks is saved for testing. In each time slot, the action of all UAVs is generated together by actor network. In our paper, as the fully-connected hidden layers are applied, the computational complexity for generating action of UAVs is $\mathcal{O}\big(\sum_{l=1}^{L}n_l\cdot n_{l-1}\big)$, where $L$ is the number of network layers, $n_l$ is the number of neurons in the $l$-th layer. Then, the computational complexity of matching algorithm is $\mathcal{O}(NM)$. The overall complexity of RAT algorithm in testing process is $\mathcal{O}\big((\sum_{l=1}^{L}n_l\cdot n_{l-1}+NM)T\big)$.}

\section{Extension to 3-D Channel Model}

\textcolor{black}{In this section, in order to consider the more practical environment and the impacts of blockage and shadowing, we extend the previous free-space to 3-D channel model proposed in~\cite{al2014optimal}. In each time slot, we assume the UAV can fly with a vertical direction $\theta^v_j(t) \in [0, \pi]$, a horizontal direction $\theta^h_j(t) \in [0,2\pi]$, and a flying distance $d_j(t) \in [0,d^{\text{max}}]$. We define the coordinate of the $j$-th UAV in the $t$-th time slot as $[X_j(t),Y_j(t),Z_j(t)]$, where $X_j(t)=X_j(0)+\sum_{l=1}^{t}d_j(l)\text{sin}\big(\theta^v_j(l)\big)\text{cos}\big(\theta^h_j(l)\big)$, $Y_j(t)=X_j(0)+\sum_{l=1}^{t}d_j(l)\text{sin}\big(\theta^v_j(l)\big)\text{sin}\big(\theta^h_j(l)\big)$, $Z_j(t)=Z_j(0)+\sum_{l=1}^{t}\text{cos}\big(\theta^v_j(l)\big)$, and $[X_j(0), Y_j(0), Z_j(0)]$ is the initial coordinate of the UAV. }
For collision avoidance, we consider
\textcolor{black}{\begin{equation}
		\begin{aligned}
			Z^{\text{min}} \leq Z_j(t) \leq Z^{\text{max}}, \forall t \in \mathcal{T},
		\end{aligned}
\end{equation}}
\textcolor{black}{where $Z^{\text{min}}$ and $Z^{\text{max}}$ are the minimal and maximal flying altitude of the UAV.}
 
\textcolor{black}{Thus, the distance between the $j$-th UAV and the $i$-th UE in $t$-th time slot is given by}
\textcolor{black}{\begin{equation}
		\begin{aligned}
			d_{ij}(t) &= \sqrt{\big(X_j(t)-x_i\big)^2+\big(Y_j(t)-x_i\big)^2+Z_j(t)},\\&\quad\quad\quad\quad\quad\quad\quad~\forall j \in \mathcal{M}, i \in \mathcal{N}, t \in \mathcal{T}.
		\end{aligned}
\end{equation}}

\textcolor{black}{The coverage radius of the $j$-th UAV in the $t$-th time slot can be given by}
\textcolor{black}{\begin{equation}
		\begin{aligned}
			R^{\text{max}}_j(t) = Z_j(t)\text{tan}(\theta^{\text{max}}).
		\end{aligned}
\end{equation}}

\textcolor{black}{The mean path loss between the $j$-th UAV and the $i$-th UE in the $t$-th time slot can be expressed as ~\cite{al2014optimal}}
\textcolor{black}{\begin{equation}
		\begin{aligned}
			L_{ij}(t)=& \frac{\eta_{\text{LoS}}-\eta_{\text{NLoS}}}{1+a \text{exp}(-b(\theta_{ij}(t)-a))}+20\text{log}_{10}\big(d_{ij}(t)\big)\\&+20\text{log}_{10}\big(\frac{4\pi f_c}{c}\big) + \eta_{\text{NLoS}},
		\end{aligned}
\end{equation}}
\textcolor{black}{where $\eta_{\text{LoS}}$, $\eta_{\text{NLoS}}$ are the path loss of achieving LoS and NLoS links, $a$ and $b$ are constant values that can be obtained in~\cite{al2014optimal}, $\theta_{ij}(t) = \text{arctan}\bigg(\frac{Z_j(t)}{R_{ij}(t)}\bigg)$ is the elevation angle between the UAV and the UE, $f_c$ is the carrier frequency, and $c$ is the light speed.} Then, we can show the data rate as follows:
\textcolor{black}{\begin{equation}
		\begin{aligned}
			r_{ij}(t) = B \text{log}_2\bigg(1+\frac{P^{\text{Tr}}}{\sigma^2}10^{-\frac{L_{ij}(t)}{10}}\bigg).
		\end{aligned}
\end{equation}}

\textcolor{black}{Additionally, we consider to maximize the energy efficiency of UAVs and motivated by~\cite{9171468}, we show the power consumed by the $j$-th UAV in the $t$-th time slot as follows}
\textcolor{black}{\begin{equation}
		\begin{aligned}
			P_j(t) =&  P_o\bigg(1+3\big(\frac{v_j(t)}{U_b}\big)^2\bigg) + P_s\bigg(\sqrt{1+\frac{1}{4}\big(\frac{v_j(t)}{V_h}\big)^4 } - \frac{1}{2}\big(\frac{v_j(t)}{V_h}\big)^2\bigg)^{\frac{1}{2}} \\&+ \frac{\pi}{2} d_0\rho_a r_s R_r^2v_j(t)^3 + wgv_j(t)\text{cos}\big(\theta^v_j(t)\big),
		\end{aligned}
\end{equation}}
\textcolor{black}{where $P_o$ and $P_s$ are fixed constants that can be obtained in~\cite{zeng2019energy}, $U_b$ is the tip speed of the rotor blade, $V_h$ denotes the mean rotor induced velocity when hovering, $d_0$ is the drag ratio of main body, $\rho_a$ is the air density, $r_s$ is the rotor solidity, $R_r$ means the rotor radius, $w$ is the weight of UAV, and $g$ is the gravity acceleration.}

\textcolor{black}{Thus, the remaining energy of the $j$-th UAV in the $t$-th time slot is defined as}
\textcolor{black}{\begin{equation}
		\begin{aligned}
			e_j(t) = e^{\text{max}}-\sum_{l=1}^{t}P_j(l)T^{\text{max}},
		\end{aligned}
	\end{equation}
	where $e^{\text{max}}$ is the maximal energy of each UAV.}

\textcolor{black}{Thus, the optimization problem can be written as follows:}
\textcolor{black}{\begin{subequations}\label{rat3d}
		\begin{IEEEeqnarray}{s,lCl'lCl'lCl}
			& \IEEEeqnarraymulticol{9}{l}{\mathcal{P}1: \underset{\bm{U},\bm{A},\bm{F}}{\text{min}}~
				\sum_{t=1}^{T}\bigg( \sum_{j=0}^{M}\sum_{i=1}^{N}a_{ij}(t)E_{ij}(t)+ k_z\sum_{j=1}^{M}P_j(t)T^{\text{max}} \bigg)} \nonumber\\&\\
			& \text{subject to:} ~(\ref{ww14}\text{b}), (\ref{ww14}\text{c}), (\ref{ww14}\text{d}), (\ref{ww14}\text{e}), (\ref{ww14}\text{f}), \nonumber\\&\quad\quad\quad(\ref{ww14}\text{g}), (\ref{ww14}\text{h}), (\ref{ww14}\text{j}), (\ref{ww14}\text{k}), \nonumber\\
			& 0 \leq \theta^v_j(t) \leq \pi,~\forall j \in \mathcal{M}, t \in \mathcal{T},\\
			& Z^{\text{min}} \leq Z_j(t) \leq Z^{\text{max}},~\forall j \in \mathcal{M}, t \in \mathcal{T},\\
			& a_{ij}(t)R_{ij}(t) \leq R^{\text{max}}_j(t),~\forall i \in \mathcal{N}, j \in \mathcal{M}, t \in \mathcal{T}.
		\end{IEEEeqnarray}
\end{subequations}}
\textcolor{black}{where $\bm{U}= \{\theta^v_j(t), \theta^h_j(t), d_j(t),~\forall j \in \mathcal{M}, t \in \mathcal{T}\}$, $k_z$ is the weight factor.}

\textcolor{black}{To solve the above problem, we define the state and action as follows:}
\textcolor{black}{\begin{itemize}
		\item[1)] \textbf{State} $s(t)$: $s(t) = \{[X_j(t), Y_j(t), Z_j(t), e_j(t)],~\forall j \in \mathcal{M}\}$.
		\item[2)] \textbf{Action} $c(t)$: the action set can be defined as $c(t) = \{[\theta^v_j(t), \theta^h_j(t), d_j(t)],~\forall j \in \mathcal{M}\}$.
		\item[3)] \textbf{Reward} $z(t)$: we define the reward as follows
		\begin{equation}
			\begin{aligned}
				z(t) = -\sum_{j=0}^{M}\sum_{i=1}^{N}a_{ij}(t)E_{ij}(t) - k_z\sum_{j=1}^{M}P_i(t)T^{\text{max}} - p,
			\end{aligned}
		\end{equation}
		where $p$ is the penalty if any of UAV flies out of the target area, i.e., if (\ref{ww14}g), (\ref{ww14}h) or (\ref{rat3d}c) is not satisfied.
\end{itemize} }

Thus, having defined the state, action and reward, the above problem can be solved by the proposed RAT algorithm as introduced before.

\section{Simulation Results}\label{section4}

\textcolor{black}{In this section, both convex optimization-based CAT and DRL-based RAT are evaluated with simulations implemented on Intel i5-3450t, NVIDIA GTX 1050Ti, Python 3.6, PULP 1.6.10, CVXPY 1.1.7, and Tensorflow 1.15.0. We deploy three fully-connected hidden layers with 1024, 800 and 600 neurons in both actor and critic networks in RAT. The actor network is trained by applying RMSPropOptimizer with the learning rate 0.001, whereas the critic network is trained by using AdamOptimzer with the learning rate 0.001. In the simulation, we assume there are 60 time slots in each training epoch. There are 100 UEs randomly distributed in a rectangle-shaped area with the side length of $X^{\text{max}} = 400$ m and $Y^{\text{max}} = 400$ m. Additionally, there are 2 UAVs deployed to serve UEs within the target area. Note that for RAT, each UAV has 20 different taking off points during the training procedure. Besides, in each time slot, UE generates a task with communication requirement $D_i(t) \in [10, 50]$ KB and computation requirement $F_i(t) \in [2\times 10^9, 2\times 10^{10}]$ cycles. Other parameters are summarized in Table~\ref{tab1}. We assume in each time slot, UAVs will send a signal to activate the corresponding UEs, which will either offload the task or execute locally, within the delay requirement.}

\begin{table}[!htpb]
	\centering
	\caption{Simulation Parameters}
	\begin{tabular}{|c|c|c|c|}
		\hline
		\textbf{Parameters} & \textbf{Settings} & \textbf{Parameters} & \textbf{Settings} \\ \hline
		\text{$T$} & \text{60} & \text{$N$} & \text{100} \\ \hline
		\text{$M$} & \text{2} & $V^{\text{max}}$ & \text{30} \\ \hline
		$d^{\text{max}}$ & \text{30} m & $T^{\text{max}}$ & \text{1} s \\ \hline
		$X^{\text{max}}$ & \text{400} m & $Y^{\text{max}}$ & \text{400} m \\ \hline
		$\theta^{\text{max}}$ & $\frac{\pi}{4}$ & $Z_j(0)$ & \text{75} m \\ \hline
		$v_i$ & \text{3} & $g_0$ & 1.42 $\times 10^{-4}$ \\ \hline
		$P^{\text{Tr}}$ & \text{0.1} W & $B$ & 10 \text{MHz}  \\ \hline
		$\sigma^2$ & \text{-90} dbm & $e^{\text{max}}$ & $10^6$ J  \\ \hline
		$k_i$ & $10^{-28}$ & $f^{\text{max}}$ & 100 GHz \\ \hline
		$\gamma$ & \text{0.999} & $p$ & \text{100} \\ \hline
		$k^{\text{max}}$ & \text{3000} & $\rho$ & \text{2} \\ \hline
		$w$ & 2 \text{kg}  & $g$ & 10 $\text{m/s}^2$ \\ \hline
		$\tau$ & 0.001 & $Z^{\text{min}}$ & 50 m \\ \hline
		$Z^{\text{max}}$ & \text{120} m & $\eta_{\text{LoS}}$ & \text{1.6} \\ \hline
		$\eta_{\text{NLoS}}$ & \text{23} & $a$ & \text{12.08} \\ \hline
		$b$ & \text{0.11} & $f_c$ & \text{2.5} \text{GHz} \\ \hline
		$c$ & \text{3}$\times 10^8$ \text{m/s} & $k_z$ & \text{0.0025} \\ \hline
		$P_o$ & \text{79.86} & $U_b$ & \text{120} m/s \\ \hline
		$P_s$ & \text{88.63} & $V_h$ & \text{4.03} \\ \hline
		$d_0$ & \text{0.6} & $\rho_a$ & \text{1.25} $\text{kg/m}^3$ \\ \hline
		$r_s$ & \text{0.05} & $R_r$ & \text{0.4} m \\ \hline
	\end{tabular}
	\label{tab1}
\end{table}

\textcolor{black}{In order to evaluate the performance of the proposed CAT and RAT, we present the following three algorithms for comparison purpose.}
\textcolor{black}{\begin{itemize}
		\item \textbf{Local Execution (LE):} All tasks are executed locally without offloading.
		\item \textbf{Random moving (RM):} In this setting, each UAV randomly selects the horizontal direction and flying distance to take.
		\item \textbf{Cluster moving (CM):} We group all the UEs into 10 clusters and each UAV flies in the trajectory connecting all the cluster center one by one. Note that it takes $\frac{T}{10}$ time slots for each UAV to move from one cluster center to another one. 
		\item \textbf{Deep Deterministic Policy Gradient (DDPG)~\cite{lillicrap2015continuous}:} We set the parameter of DDPG the same as actor and critic networks of RAT, but do not apply the prioritized experience replay. In other words, DDPG uniformly samples the experiences from the experience replay buffer in the training procedure.
	\end{itemize}
	Note that both RM, CM, DDPG apply the matching algorithm proposed in Algorithm~\ref{match} to decide the user association and resource allocation. }

\subsection{Convergence Evaluation of CAT and RAT}

\begin{figure}[htpb]
	\centering
	\includegraphics[width=3in]{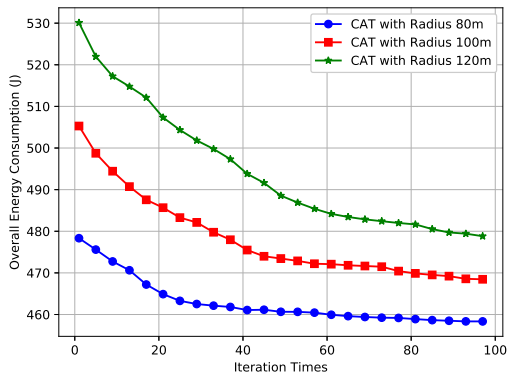}
	\caption{The convergence performance of proposed CAT.}\label{cvxfig}
\end{figure}

\textcolor{black}{In this subsection, we show the convergence of proposed CAT and RAT. In Fig.~\ref{cvxfig}, we depict the convergence performance of CAT with three different pairs of initial trajectories. Specifically, we group all UEs into one cluster and the UAVs fly in a circle around the cluster center with radius $80$ m, $100$ m, and $120$ m respectively. We denote these three pairs of UAV trajectories as the initial trajectories. As shown in Fig.~\ref{cvxfig}, we can conclude that for any initial trajectory, the overall energy consumption of UEs achieved by CAT always decreases and finally remains stable after several iteration times. However, one can also observe that the convergent solution achieved by CAT will be influenced by the initial trajectory.}

\begin{figure}[htpb]
	\centering
	\subfigure[The overall energy consumption of RAT with different batch size.]{
		\label{rat_batch} 
		\includegraphics[width=1.6in]{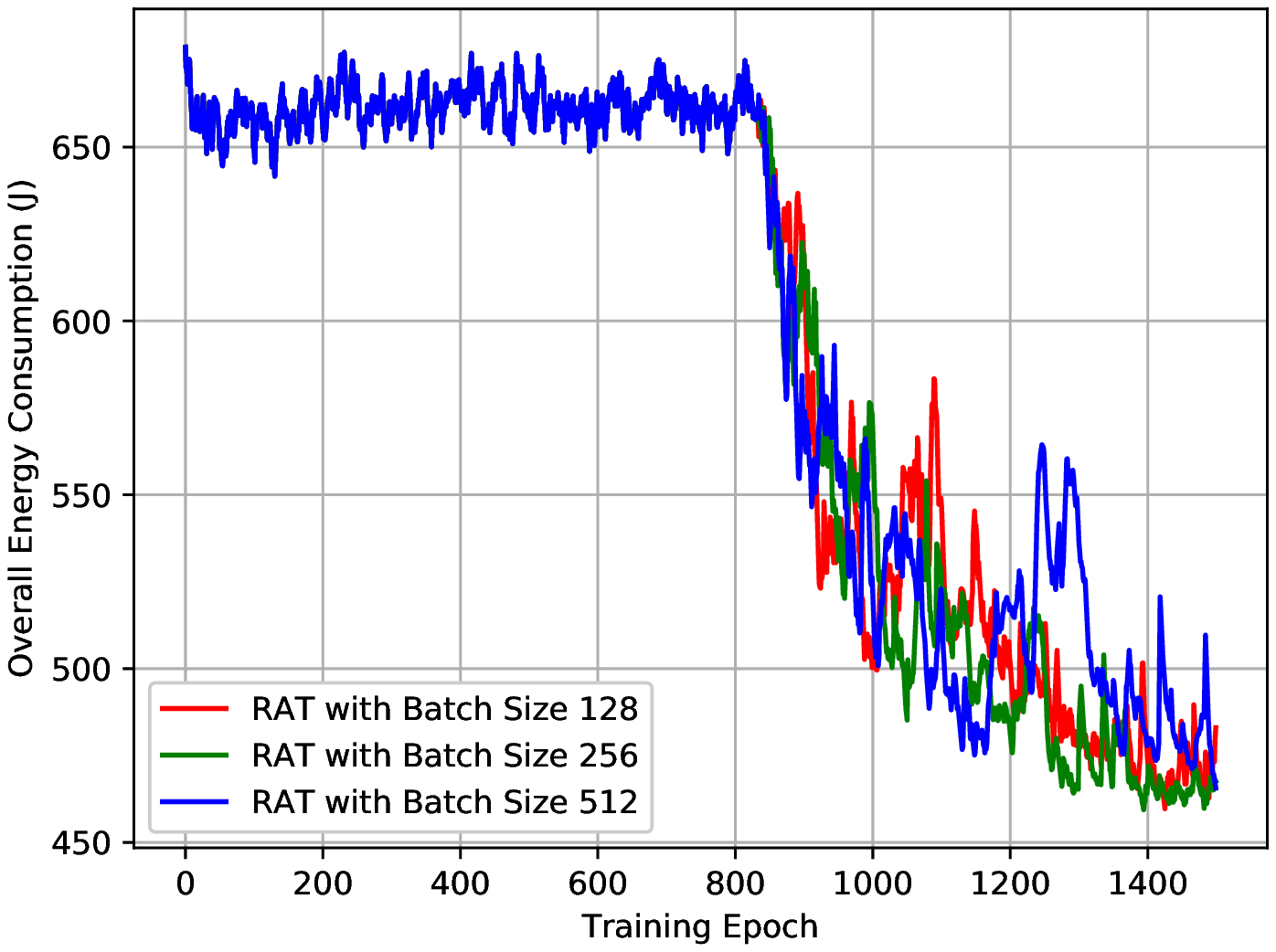}}
	\hspace{0.1in}
	\subfigure[The overall energy consumption of DDPG with different batch size.]{
		\label{um_batch} 
		\includegraphics[width=1.6in]{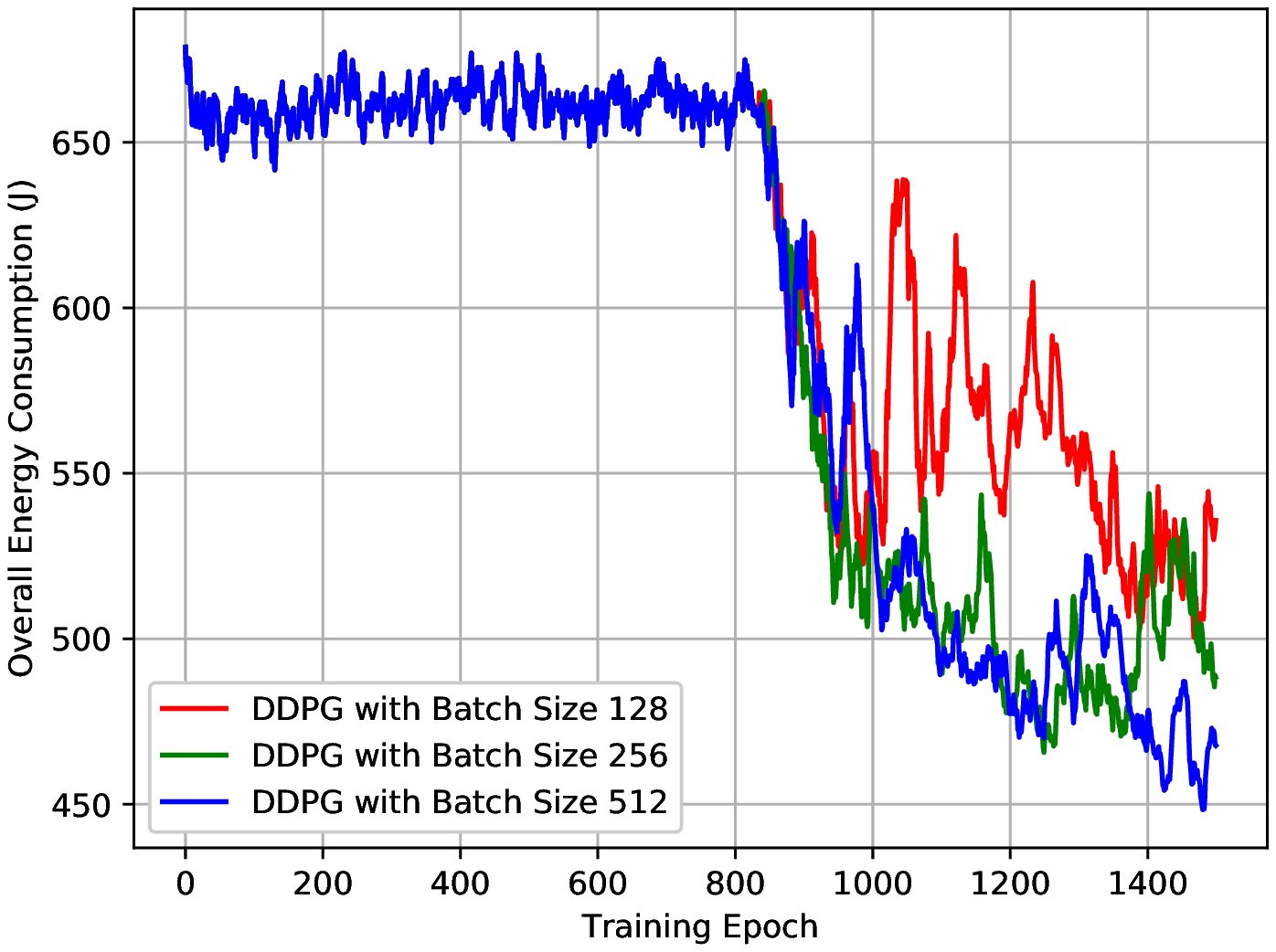}}
	\caption{The convergence performance of RAT and DDPG with different size of mini-batch.}
	\label{batch_comp} 
\end{figure} 

\begin{figure}[htpb]
	\centering
	\subfigure[The overall energy consumption of RAT with different buffer size.]{
		\label{rat_buffer} 
		\includegraphics[width=1.6in]{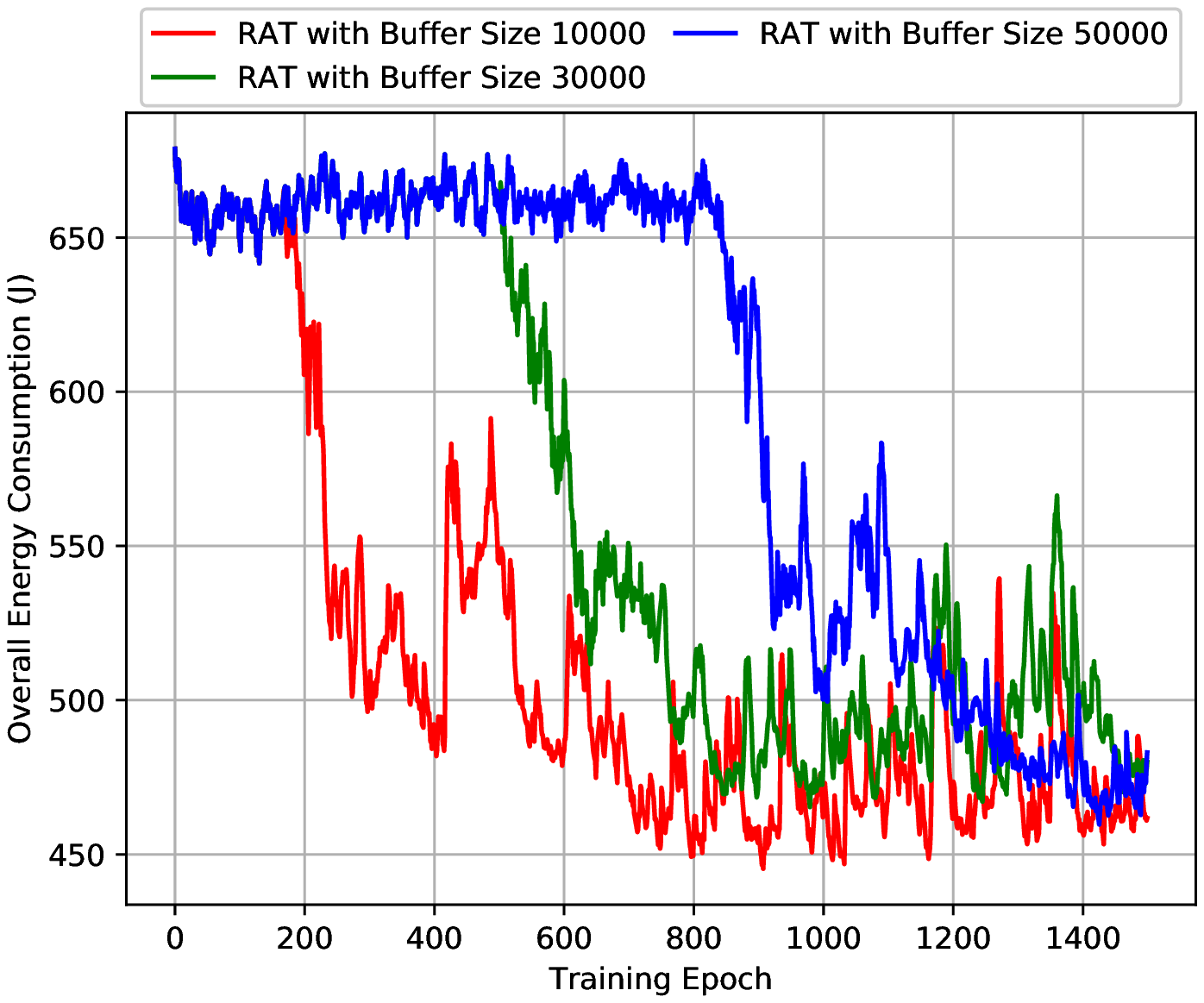}}
	\hspace{0.1in}
	\subfigure[The overall energy consumption of DDPG with different buffer size.]{
		\label{um_buffer} 
		\includegraphics[width=1.6in]{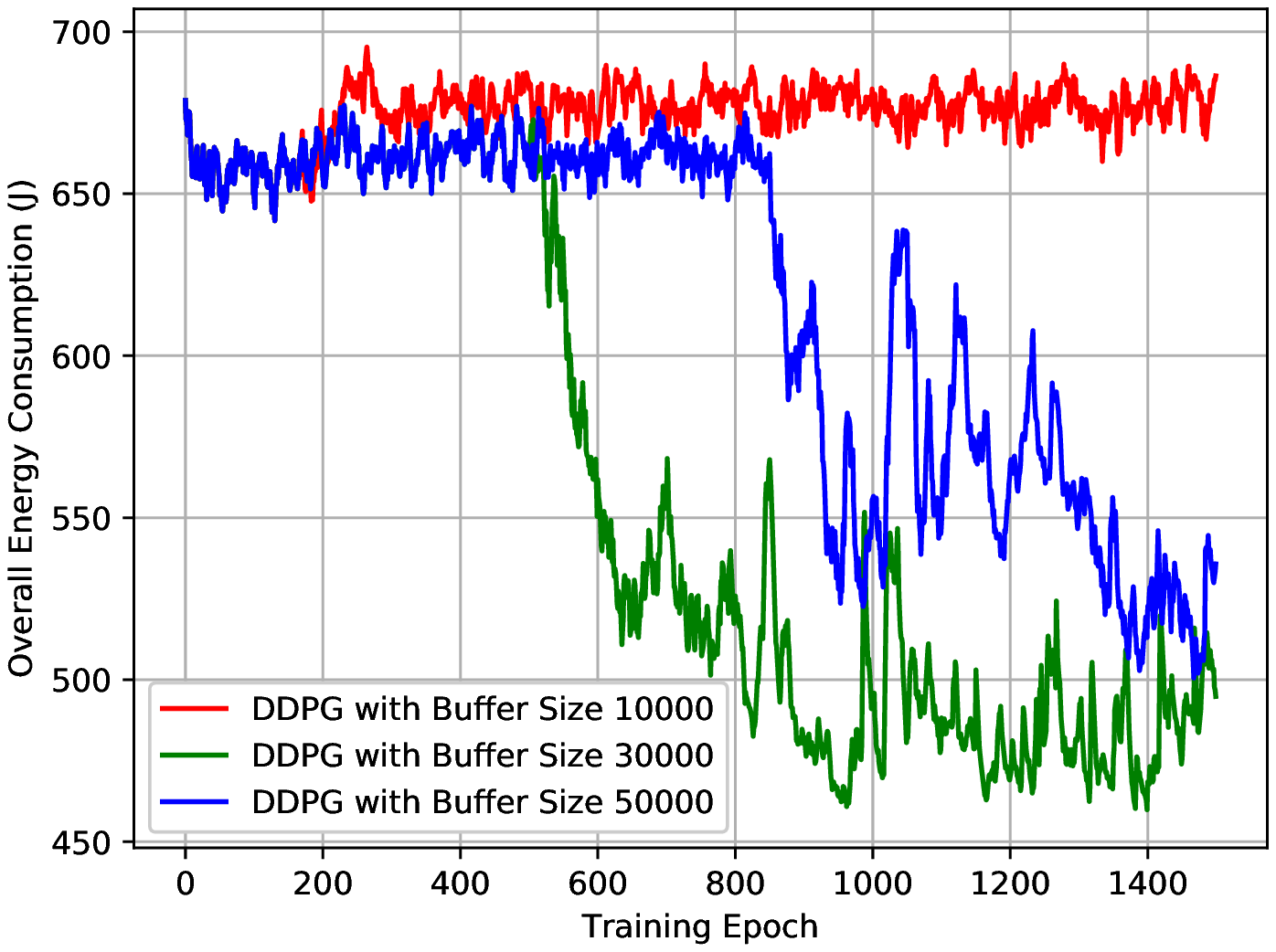}}
	\caption{The convergence performance of RAT and DDPG with different experience replay buffer.}
	\label{buffer_comp} 
\end{figure} 

\textcolor{black}{Then, we show the convergence performance of RAT in training process. From Fig.~\ref{batch_comp} to Fig.~\ref{buffer_comp}, we compare the influence of hyperparameters to both DDPG and RAT. Prioritized experience replay is applied in RAT. Both RAT and DDPG start the learning procedure once the experience replay buffer is full. In Fig.~\ref{batch_comp}, we depict the overall energy consumption of RAT and DDPG for different size of mini-batches, where the size of experience replay buffer is $50000$. To be more specific, from Fig.~\ref{rat_batch}, we can see that RAT has the similar convergence performance for different size of mini-batches and it becomes more stable during the learning procedure. In Fig.~\ref{um_batch}, when the batch size is 128, DDPG has an obvious fluctuation during the learning procedure. When the batch size is 256, the convergence performance of DDPG becomes worse after the 1400-th epoch. While DDPG can only have a promising convergence performance when the batch size is $512$. Overall, from Fig.~\ref{batch_comp}, it is clear to see that the RAT is less sensitive to the change of mini-batch than DDPG. }

\textcolor{black}{In Fig.~\ref{buffer_comp}, we depict the overall energy consumption of RAT and DDPG for different sizes of experience replay buffer, where the size of mini-batch is set as 128. From Fig.~\ref{rat_buffer} and~\ref{um_buffer}, when the buffer size is $10000$, the proposed RAT finally remains stable between 450 $\text{J}$ and 500 $\text{J}$, although it has an obvious fluctuation during the learning process. The DDPG has no convergence tendency during the entire learning procedure. When the buffer size is $50000$, DDPG becomes worse after $1000$-th epoch, and finally reaches 550 $\text{J}$. Overall, we can observe that DDPG can only have a promising performance when the buffer size is $30000$, while RAT can always converge and remain stable during the learning procedure, no matter which the buffer size is. Thus, we can conclude that RAT is less sensitive to the size of experience replay buffer than DDPG.}

\subsection{Trajectory Evaluation of CAT and RAT}

\begin{figure}[htpb]
	\centering
	\includegraphics[width=3in]{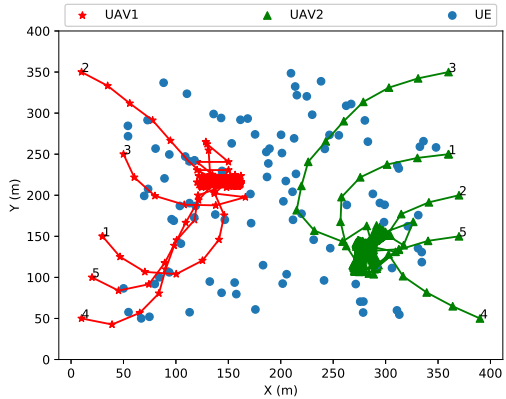}
	\caption{Multi-UAV enabled F-MEC controlled by RAT.}\label{uav2_move}
\end{figure}

\begin{figure}[!t]
	\centering
	\includegraphics[width=3in]{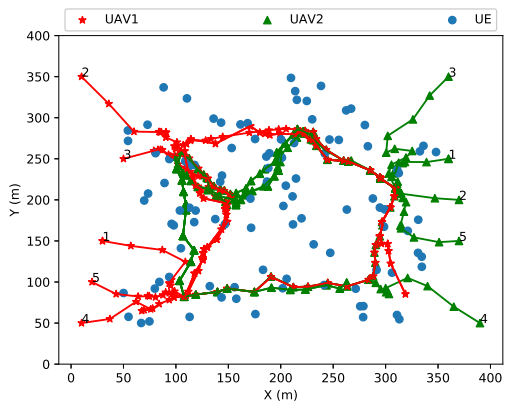}
	\caption{Multi-UAV enabled F-MEC controlled by CAT.}\label{cat_move}
\end{figure}

\textcolor{black}{In Fig.~\ref{uav2_move} and Fig.~\ref{cat_move}, we show the trajectories obtained by RAT and CAT, respectively. Note that during the training procedure, the UAVs controlled by RAT always starts to serve UEs from 20 different taking off points. Additionally, for fairness, the UAVs controlled by CAT have the same taking off points as RAT. For the initial trajectories, we group all the UEs into 6 clusters and each UAV flies in the trajectory connecting all cluster centers one by one. Note that the iteration number of CAT is 10.}

\textcolor{black}{As shown in Fig.~\ref{uav2_move}, we randomly select 5 pairs of taking off points for comparison. One can observe that no matter which the taking off points of the UAVs are, the proposed RAT can guide the UAVs to their certain areas and move around to serve different UEs. This is due to the fact that we train the RAT to converge during the training stage by randomly generating several taking off points of the UAVs. Then, during the testing stage, RAT can intermediately output the best solutions once taking off points are given. }

\textcolor{black}{In Fig.~\ref{cat_move}, one can also see that the trajectories obtained by CAT are similar with the initial trajectories. This may indicate that CAT may fall into the local optimum, whereas the proposed RAT has the global search ability due to the exploration feature of DRL.}

\subsection{Energy Consumption Evaluation of CAT and RAT}

\begin{figure}[htpb]
	\centering
	\subfigure[The overall energy consumption of RAT, CAT, RM, CM, LE with different taking off points.]{
		\label{ene_eps} 
		\includegraphics[width=1.6in]{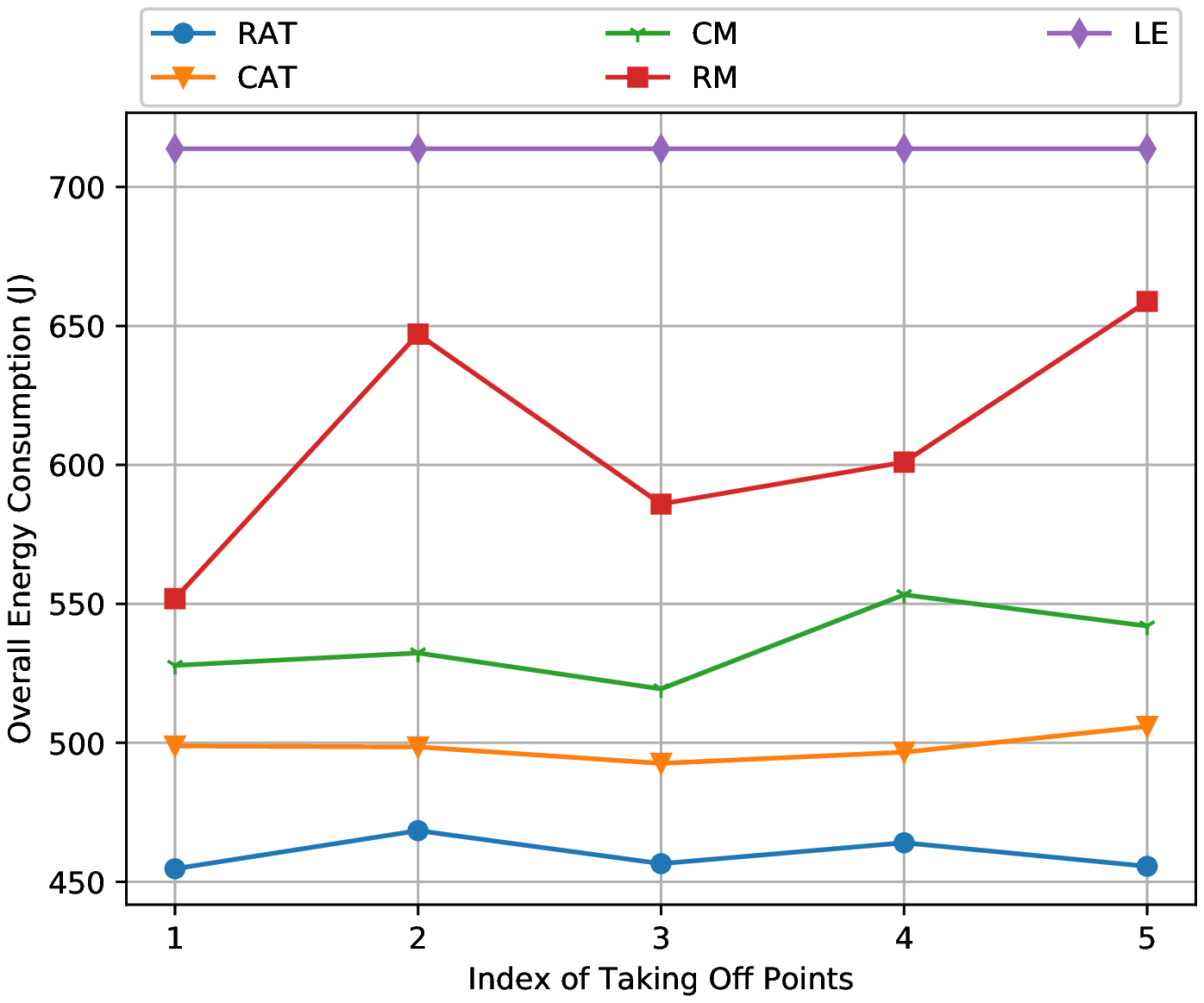}}
	\hspace{0.1in}
	\subfigure[The overall energy consumption of RAT, CAT, RM, CM, LE in different number of time slots.]{
		\label{ene_step} 
		\includegraphics[width=1.6in]{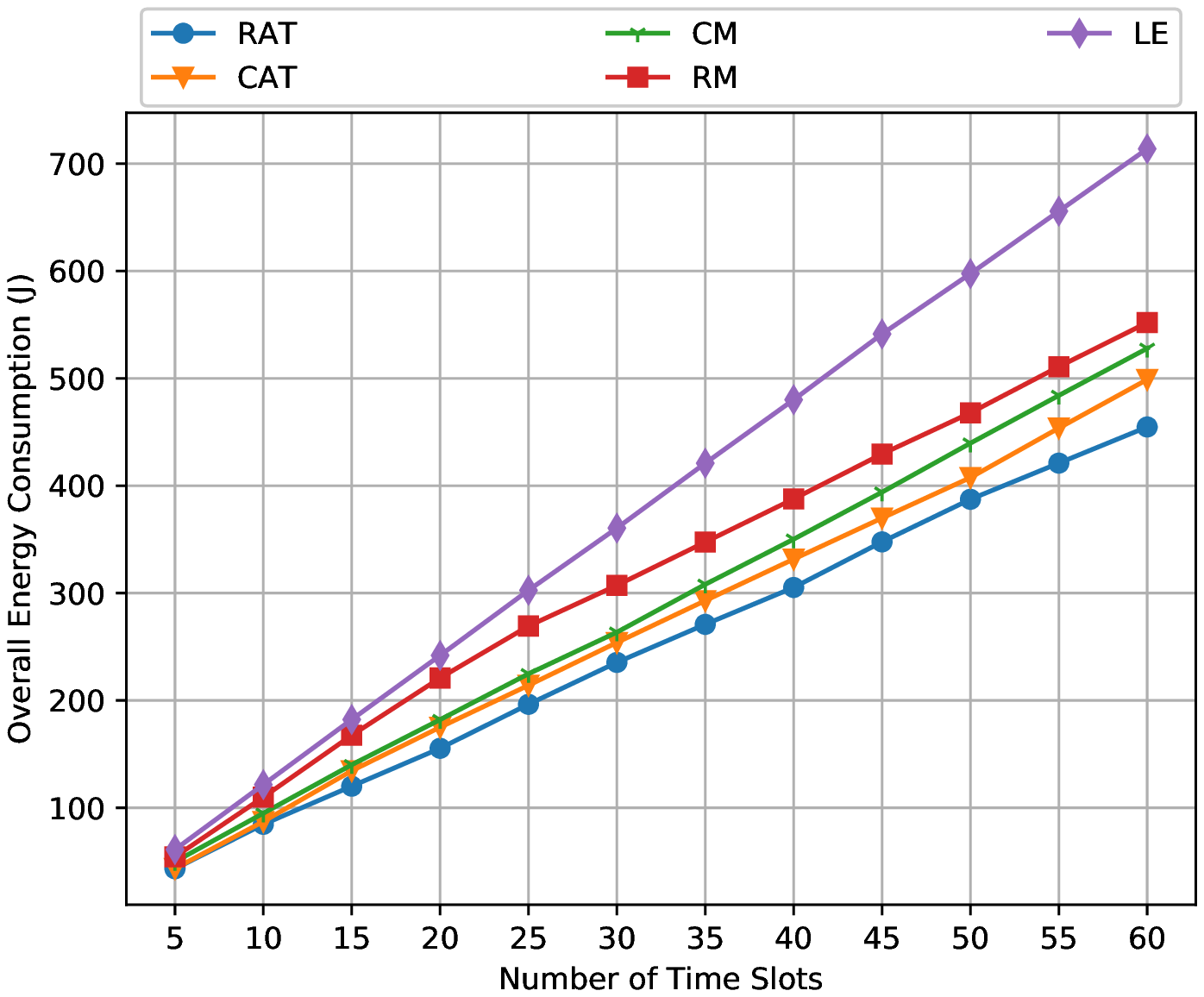}}
	\caption{The performance comparison of RAT, CAT, RM, CM, and LE.}
	\label{compare_2d} 
\end{figure} 

\textcolor{black}{In Fig.~\ref{compare_2d}, we compare the performance of RAT, CAT, CM, RM and LE in terms of energy consumption of UEs. As shown in Fig.~\ref{compare_2d} (a), we depict the overall energy consumption of UEs achieved by RAT, CAT, CM, RM, and LE with different taking off points. It is obvious to see that LE has the worst performance. This is because all UEs execute their tasks locally without offloading, which will inevitably consume more energy. RM outperforms LE but it fluctuates with the index of taking off points. CM has better performance than RM, which always remains between 520 J and 550 J. CAT outperforms LE, RM, and CM, which remains about 500 J. Additionally, one can observe that RAT achieves the best performance, as expected.}

\textcolor{black}{Furthermore, we depict the overall energy consumption of UEs achieved by RAT, CAT, RM, CM, and LE in different number of time slots in Fig.~\ref{compare_2d} (b), with the index of taking off points setting as 1. It is readily to see that both the energy consumption of RAT, CAT, RM, CM, and LE increase as the number of time slots increases. LE performs the worst, which consumes above 700 J eventually. Additionally, we can observe that RAT outperforms other algorithms. Moreover, CAT still has considerable performance, which is only slightly worse than RAT.}

\begin{table}[!htpb]\label{T:equipos}
	\centering
	\caption{Executed Time of CAT and RAT}
	\begin{center}
		\begin{tabular}{| c | c | c | c |}
			\hline
			\multirow{2}{*}{\textbf{Index} }& \multirow{2}{*}{\textbf{CAT} (s)} & \multicolumn{2}{ c |}{\textbf{RAT}}  \\ 
			\cline{3-4}
			&  & \textbf{Training} (s) & \textbf{Testing} (s)  \\
			\hline
			1 & 1405.23 & \multirow{5}{*}{10534.88} & 1.23  \\ 
			2 & 1491.74 &  & 1.22  \\ 
			3 & 1460.46 &  & 1.20 \\ 
			4 & 1445.11 &  & 1.21 \\ 
			5 & 1402.48 &  & 1.21  \\ \hline
		\end{tabular}
	\end{center}
	\label{tabt}
\end{table}

\textcolor{black}{In Table~\ref{tabt}, we show the time consumed by CAT and RAT for each pair of taking off points in Fig.~\ref{compare_2d}. Note that RAT is trained for 3000 epochs, while the iteration number of CAT is 10. One can see that for all the taking off points, the proposed CAT takes over 1400 seconds to find solutions, while RAT only takes 1.2 seconds in average, although it takes longer time in training process. This is because once the RAT are trained properly, it only needs a few number of algebra calculations to obtain the solution.}

\begin{figure}[!htpb]
	\centering
	\includegraphics[width=3in]{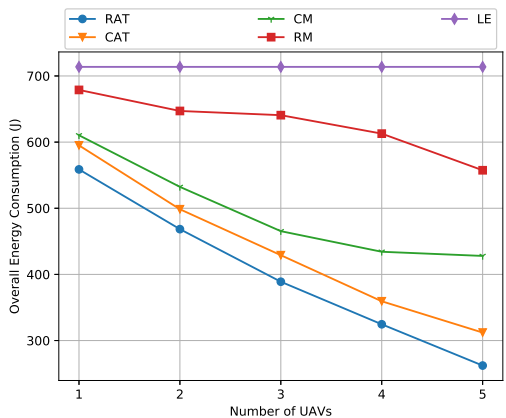}
	\caption{\textcolor{black}{The overall energy consumption of RAT, CAT, RM, CM, LE with different number of UAVs.}}\label{num_uav}
\end{figure}

\textcolor{black}{Additionally, in Fig.~\ref{num_uav}, we analyse the overall energy consumption of RAT, CAT, RM, CM and LE when we have different number of UAVs. Note that for fairness, the UAVs controlled by RAT, CAT, RM, CM have the same taking off points. Specifically, in Fig.~\ref{num_uav}, one observes that the energy consumption of UEs achieved by RAT, CAT, RM, and CM decrease with the increasing number of UAVs. This is because deploying more UAVs provides higher computational capacity. Therefore, more UEs will benefit from offloading, which will decrease their overall energy consumption. Besides, we observe that for all the cases, RAT can achieve the best performance, whereas CAT performs slightly worse than RAT. Also, CM, LM and RM have worse performance than CAT, as expected.}

\subsection{Extension to 3-D channel model}

\begin{figure}[!htpb]
	\centering
	\includegraphics[width=3in]{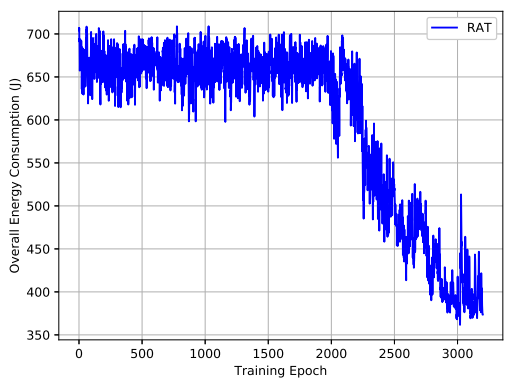}
	\caption{The convergence performance of proposed RAT in 3-D UAV trajectory and 3-D channel model scenario.}\label{train3d}
\end{figure}

\textcolor{black}{In this subsection, we analyse the performance of proposed RAT in 3-D channel model. We set the number of time slots $T$ as 50, the channel bandwidth as 20 $\text{MHz}$,  $D_i(t) \in [5, 10]$ $\text{KB}$, $F_i(t) \in [7.5\times 10^8, 2\times 10^9]$ cycles, the size of mini-batch is 512, and the size of experience replay buffer is 100000. In each training epoch, each UAV starts to serve UEs with the altitude of $Z_j(0) = $ 50 m. Firstly, we depict the overall energy consumption achieved by the proposed RAT algorithm during the training procedure in Fig.~\ref{train3d}. One can see that the overall energy consumption of UEs remains between 600 J and 700 J in the beginning. When the learning process starts, the curve decreases and eventually remains slightly above 350 J.}

\begin{figure}[!htpb]
	\centering
	\includegraphics[width=3.4in]{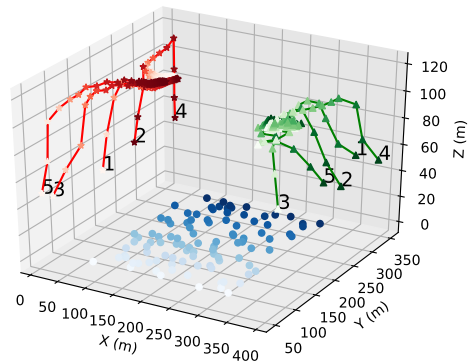}
	\caption{3-D trajectories obtained by RAT in 3-D scenario (blue dots for UEs, red stars for UAV1 and green triangles for UAV2).}\label{trajectory3d}
\end{figure}

\textcolor{black}{Then, we depict the UAV trajectories obtained by RAT during testing phase in Fig.~\ref{trajectory3d}. Note that blue dots represent UEs, red stars represent the trajectories of UAV1 and green triangles represent the trajectories of UAV2. As shown in Fig.~\ref{trajectory3d}, one can see that the UAVs always move from their taking off points to the certain areas, and move around to serve different UEs with the most sufficient distance. In addition, one can observe that each UAV will increase its altitude at the beginning. This is because higher altitude may increase the coverage radius of the UAV, thereby serving more UEs, although it also decreases the data rate of the offloading process.}

\begin{figure}[htpb]
	\centering
	\subfigure[The overall energy consumption of UEs achieved by RAT, CM, and RM with different taking off points.]{
		\label{ene3d} 
		\includegraphics[width=1.6in]{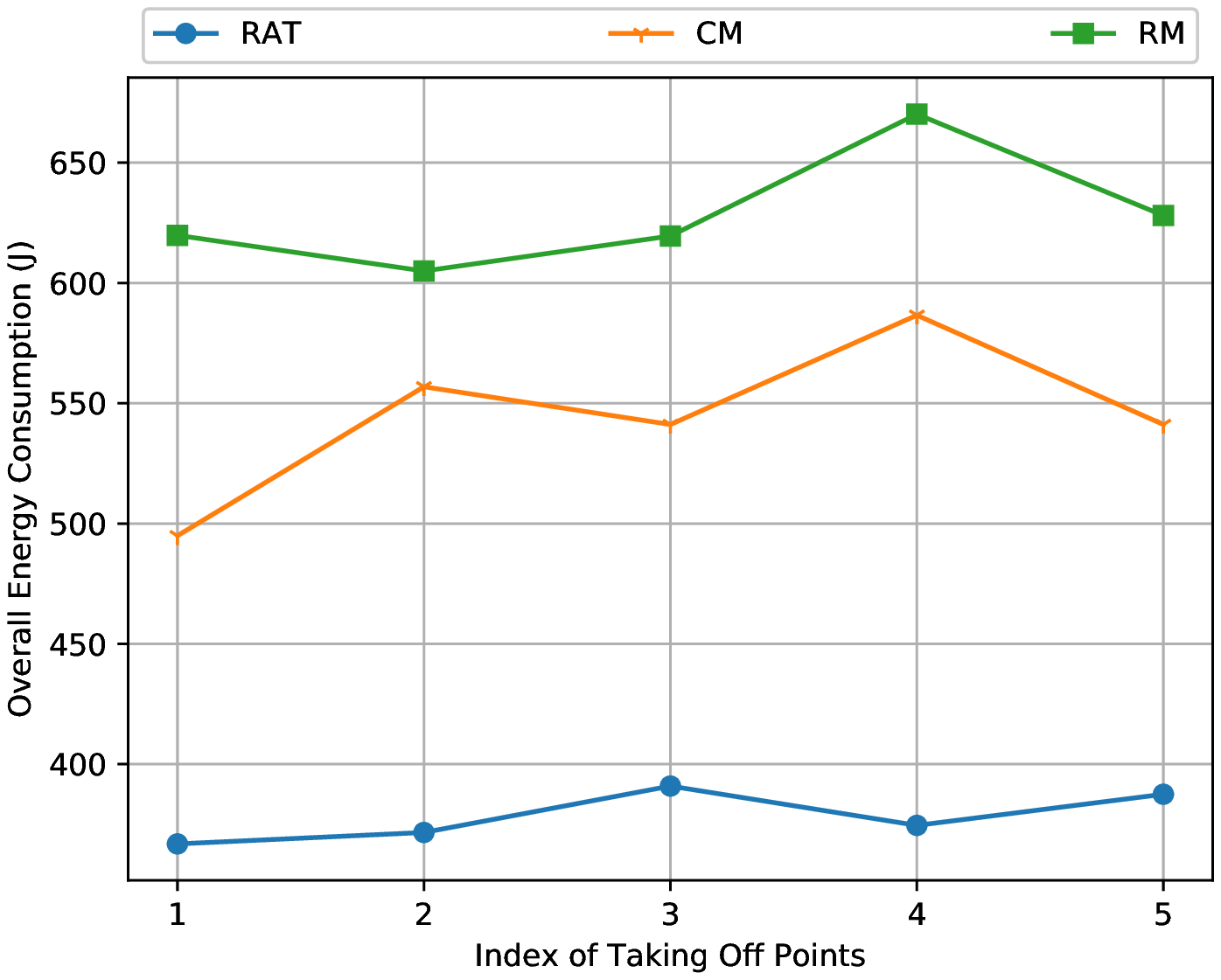}}
	\hspace{0.1in}
	\subfigure[The overall energy consumption of UAVs achieved by RAT, CM, and RM with different taking off points.]{
		\label{uene3deps} 
		\includegraphics[width=1.6in]{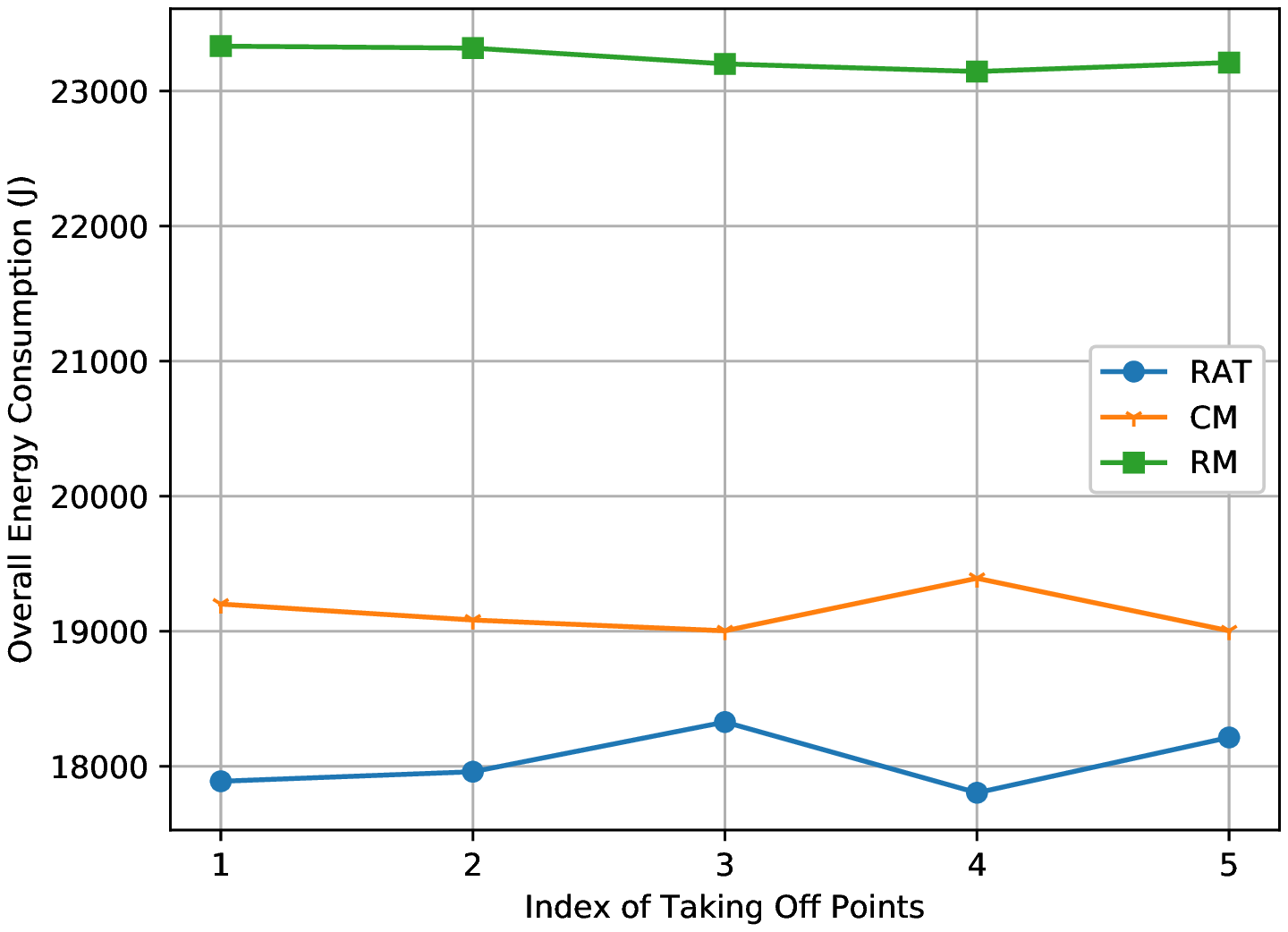}}
	\caption{The performance comparison of RAT, CM, and RM.}
	\label{compare3d} 
\end{figure}

\textcolor{black}{Furthermore, we analyse the overall energy consumption of UEs and UAVs achieved by RAT, CM, and RM in different scenarios in Fig.~\ref{compare3d}, where the UAVs controlled by CM first climb from the minimal altitude $Z^{\text{min}}$ to the maximal altitude $Z^{\text{max}}$ in the first 10 time slots, and after that fly horizontally. Also, the RM randomly selects the available flying action for each UAV, including the horizontal flying direction, the vertical flying direction, and the flying distance. More precisely, in Fig.~\ref{compare3d} (a), one can observe that our proposed RAT consistently outperforms CM and RM, whereas CM performs worse than RAT but better than RM, as expected.}

\textcolor{black}{Finally, we show the overall energy consumption of UAVs achieved by RAT, CM and RM in Fig.~\ref{compare3d} (b). One observes that our proposed RAT has the best performance, whereas CM has the worse performance than RAT, but better than RM.}

\section{Conclusion}\label{section5}
In this paper, we have considered the flying mobile edge computing architecture, by taking advantage of the UAVs to serve as the moving platform. We aim to minimize the energy consumption of all the UEs by optimizing the UAVs' trajectories, user associations and resource allocation. To tackle the multi-UAVs' trajectories problem, a convex optimization-based CAT has been first proposed. Then, in order to conduct fast decision, a DRL-based RAT including a matching algorithm has also been proposed. Simulation results show that CAT and RAT have considerable performance.

\bibliographystyle{ieeetran}
\bibliography{reference}

\begin{IEEEbiography}[{\includegraphics[width=1in,height=1.25in,clip,keepaspectratio]{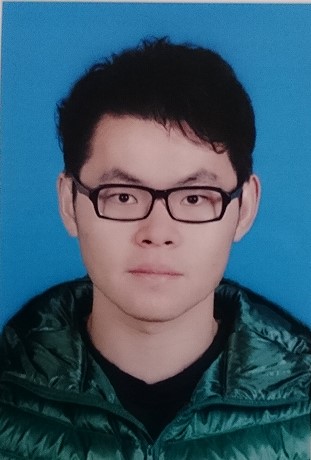}}]{Liang Wang}
	received his B.Eng. degree in 2014 and MSc. degree in 2015. He is currently working towards the Ph.D. degree in computer science with Northumbria University, Newcastle upon Tyne, U.K. His research interests include UAV communication, mobile edge computing, and machine learning.
\end{IEEEbiography}

\begin{IEEEbiography}[{\includegraphics[width=1in,height=1.25in,clip,keepaspectratio]{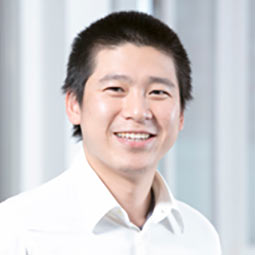}}]{Kezhi Wang}
	received his B.E. and M.E. degrees in School of Automation from Chongqing University, China, in 2008 and 2011, respectively. He received his Ph.D. degree in Engineering from the University of Warwick, U.K. in 2015. He was a senior research officer in University of Essex, U.K. from 2015-2017. Currently he is a Senior Lecturer with Department of Computer and Information Sciences at Northumbria University, U.K. 
	His research interests include mobile edge computing, intelligent reflection surface (IRS) and machine learning.
\end{IEEEbiography}

\begin{IEEEbiography}[{\includegraphics[width=1in,height=1.25in,clip,keepaspectratio]{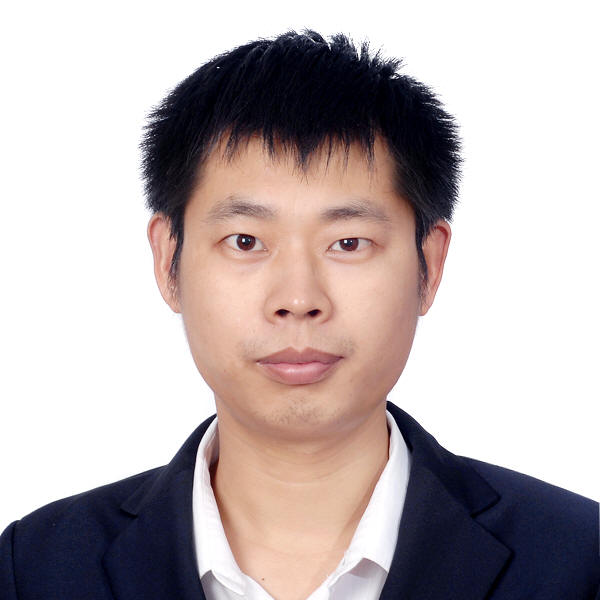}}]{Cunhua Pan}
received the B.S. and Ph.D. degrees from the School of Information Science and Engineering, Southeast University, Nanjing, China, in 2010 and 2015, respectively. From 2015 to 2016, he was a Research Associate at the University of Kent, U.K. He held a post-doctoral position at Queen Mary University of London, U.K., from 2016 and 2019, where he is currently a Lecturer. 
	
His research interests mainly include  reconfigurable intelligent surfaces (RIS), intelligent reflection surface (IRS), ultra-reliable low latency communication (URLLC) , machine learning, UAV, Internet of Things, and mobile edge computing. He serves as a TPC member for numerous conferences, such as ICC and GLOBECOM, and the Student Travel Grant Chair for ICC 2019. He is currently an Editor of IEEE Wireless Communication Letters, IEEE Communications Letters and IEEE ACCESS. He also serves as a lead guest editor of IEEE Journal of Selected Topics in Signal Processing (JSTSP)   Special Issue on Advanced Signal Processing for Reconfigurable Intelligent Surface-aided 6G Networks, lead guest editor of IEEE ACCESS Special Issue on Reconfigurable Intelligent Surface Aided Communications for 6G and Beyond.
	
\end{IEEEbiography}

\begin{IEEEbiography}[{\includegraphics[width=1in,height=1.25in,keepaspectratio]{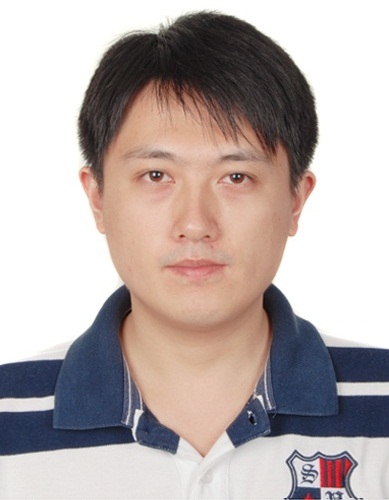}}]{Wei Xu}(S'07-M'09-SM'15) received his B.Sc. degree in electrical engineering and his M.S. and Ph.D. degrees in communication and information engineering from Southeast University, Nanjing, China in 2003, 2006, and 2009, respectively. Between 2009 and 2010, he was a Post-Doctoral Research Fellow with the Department of Electrical and Computer Engineering, University of Victoria, Canada. He is currently a Professor at the National Mobile Communications Research Laboratory, Southeast University. He is also an Adjunct Professor of the University of Victoria in Canada, and a Distinguished Visiting Fellow of the Royal Academy of Engineering, U.K. He has co-authored over 100 refereed journal papers in addition to 36 domestic patents and four US patents granted. His research interests include information theory, signal processing and machine learning for wireless communications. He was an Editor of \textsc{IEEE Communications Letters} from 2012 to 2017. He is currently an Editor of \textsc{IEEE Transactions on Communications} and an Senior Editor of \textsc{IEEE Communications Letters}. He received the Best Paper Awards from a number of prestigious IEEE conferences including IEEE Globecom/ICCC etc. He received the Youth Science and Technology Award of China Institute of Communications in 2018.
\end{IEEEbiography} 

\begin{IEEEbiography}[{\includegraphics[width=1in,height=1.25in,clip,keepaspectratio]{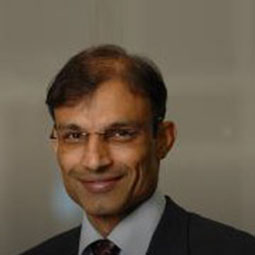}}]{Nauman Aslam} received the Ph.D. degree in engineering mathematics from Dalhousie University, Halifax, NS, Canada, in 2008. He is currently an Associate Professor with the Department of Computer Science and Digital Technologies, Northumbria University, Newcastle upon Tyne, U.K. He is also an Adjunct Assistant Professor with Dalhousie University. Prior to joining Northumbria University, he was an Assistant Professor with Dalhousie University. His research interests include wireless sensor network, energy efficiency, security, and WSN health applications.
\end{IEEEbiography}

\begin{IEEEbiography}[{\includegraphics[width=1in,height=1.25in,clip,keepaspectratio]{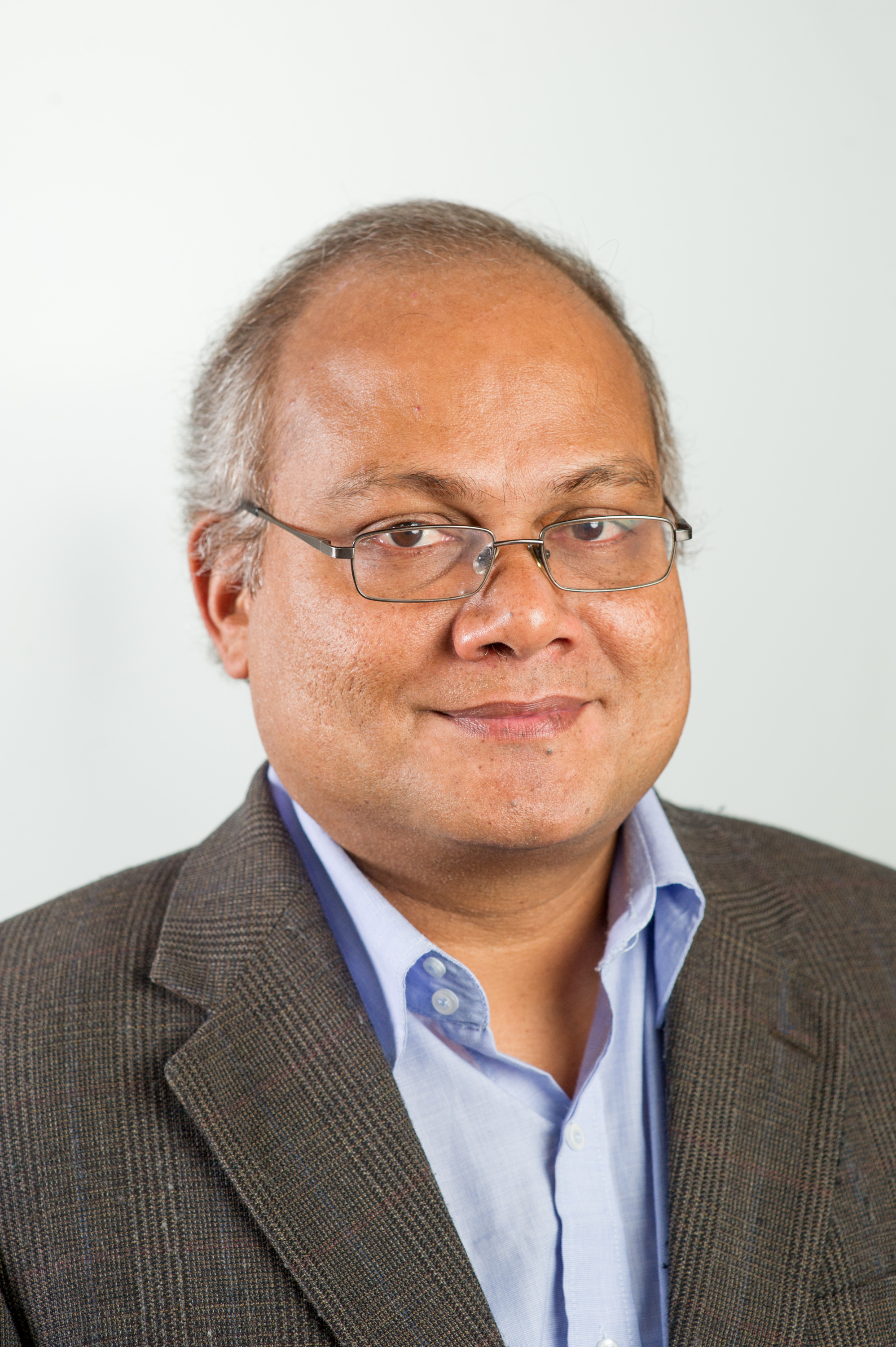}}]{Arumugam Nallanathan}(S'97-M'00-SM'05-F'17) is Professor of Wireless Communications and Head of the Communication Systems Research   (CSR) group in the School of Electronic Engineering and Computer  Science at Queen Mary University of London since September 2017. He was with the Department of Informatics at King's College London from December 2007 to August 2017, where he was Professor  of  Wireless Communications from April 2013 to August 2017 and a Visiting Professor from September 2017. He was an Assistant  Professor in the Department of Electrical and Computer Engineering, National University of Singapore from August 2000  to December 2007. His research interests include  Artificial  Intelligence for Wireless Systems, Beyond 5G Wireless Networks,  Internet of Things (IoT) and Molecular Communications. He published nearly 500 technical papers in scientific journals and international conferences. He is a co-recipient of the Best Paper Awards presented at the IEEE International Conference on  Communications 2016 (ICC'2016), IEEE Global Communications  Conference 2017 (GLOBECOM'2017) and IEEE Vehicular Technology  Conference 2018 (VTC'2018). He is an IEEE Distinguished Lecturer. He has been selected as a Web of Science Highly Cited Researcher in 2016.
	
He is an Editor for IEEE Transactions on Communications and  Senior Editor for IEEE Wireless Communications Letters. He was an Editor for IEEE Transactions on Wireless Communications  (2006-2011), IEEE Transactions on Vehicular Technology (2006-2017) and IEEE Signal Processing Letters. He served as the Chair for the Signal Processing and Communication Electronics Technical Committee of IEEE Communications Society and Technical Program  Chair and member of Technical Program Committees in numerous IEEE conferences. He received the IEEE Communications Society SPCE outstanding service award 2012 and IEEE Communications Society RCC outstanding service award 2014.
\end{IEEEbiography}

\end{document}